\documentclass[%
 reprint,
 amsmath,amssymb,
 aps,
]{revtex4-2}

\usepackage{graphicx}% Include figure files
\usepackage{amsmath}
\usepackage{pifont}
\usepackage{amssymb}
\usepackage{float}
\usepackage{xcolor}
\usepackage{mathtools}
\usepackage{tikz-cd}
\usepackage{MnSymbol}
\usepackage{xcolor}
\usetikzlibrary{calc}
\usetikzlibrary{fadings}
\usepackage{physics}
\usepackage{verbatim}
\usepackage{siunitx}
\usetikzlibrary{matrix}
\usepackage{dcolumn}% Align table columns on decimal point
\usepackage{bm}% bold math
\usetikzlibrary{decorations.markings}
\usepackage{tikz}
\usetikzlibrary{calc}
\usetikzlibrary{positioning}
\usepackage{comment}

\usepackage{graphicx}% Include figure files
\usepackage{dcolumn}% Align table columns on decimal point
\usepackage{bm}% bold math
\newcommand{\physicalpendulum}{
    \pgfpathmoveto{\pgfpoint{0cm}{2cm}}
    \pgfpathcurveto{\pgfpoint{.2cm}{2cm}}{\pgfpoint{.2cm}{2cm}}{\pgfpoint{1cm}{2cm}}
    \pgfpathcurveto{\pgfpoint{8.75cm}{0cm}}{\pgfpoint{1.8cm}{-2cm}}{\pgfpoint{2cm}{-2cm}}
    \pgfpathcurveto{\pgfpoint{-3cm}{-2cm}}{\pgfpoint{-1cm}{-2cm}}{\pgfpoint{-3cm}{-1cm}}
    \pgfpathcurveto{\pgfpoint{-4.2cm}{0cm}}{\pgfpoint{-3.5cm}{2cm}}{\pgfpoint{-1cm}{2cm}}
    \pgfsetfillopacity{0.3}
    \pgfsetfillcolor{gray!50!white}
    \pgfusepath{fill}
    }

\begin{document}

\preprint{APS/123-QED}

\title{Dynamics and Geometry of Entanglement in Many-Body Quantum Systems}% Force line breaks with \\
\author{Peyman Azodi}
\email{pazodi@princeton.edu}
 %\altaffiliation[Also at ]{Physics Department, XYZ University.}%Lines break automatically or can be forced with \\
\author{Herschel A.Rabitz}%

\affiliation{%
 Department of Chemistry, Princeton University, Princeton, New Jersey 08544, USA %\textbackslash\textbackslash
}%

\date{\today}% It is always \today, today,
             %  but any date may be explicitly specified

\begin{abstract}

A new framework is formulated to study entanglement dynamics in many-body quantum systems along with an associated geometric description. In this formulation, called the Quantum Correlation Transfer Function (QCTF), the system's wave function or density matrix is transformed into a new space of complex functions with isolated singularities. Accordingly, entanglement dynamics is encoded in specific residues of the QCTF, and importantly, the explicit evaluation of the system's time dependence is avoided. Notably, the QCTF formulation allows for various algebraic simplifications and approximations to address the normally encountered complications due to the exponential growth of the many-body Hilbert space with the number of bodies. These simplifications are facilitated through considering the \textit{patterns}, in lieu of the elements, lying within the system's state. Consequently, a main finding of this paper is the exterior (Grassmannian) algebraic expression of many-body entanglement as the collective areas of regions in the Hilbert space spanned by pairs of projections of the wave function onto an arbitrary basis. This latter geometric measure is shown to be equivalent to the second-order R\'enyi entropy. Additionally, the geometric description of the QCTF shows that characterizing features of the reduced density matrix can be related to experimentally observable quantities. The QCTF-based geometric description offers the prospect of theoretically revealing aspects of many-body entanglement, by drawing on the vast scope of methods from geometry. 
%It is shown that, when examining entanglement, this simplification is partly due to redundancies in the system's wave function.

\end{abstract}

%\keywords{Suggested keywords}%Use showkeys class option if keyword
                              %display desired
\maketitle

%\tableofcontents

\section{Introduction}\label{secp}

Entanglement, the unique and intriguing feature of the quantum realm, has been subject to extensive theoretical and experimental studies in a variety of fields. Entanglement is the cornerstone of quantum information science \cite{steane1998quantum, plenio2007introduction, PhysRevLett.96.110404,RevModPhys.81.865} while also forming a basis to explain quantum thermalization \cite{kaufman2016quantum,RevModPhys.91.021001, eisert2015quantum, sala2020ergodicity,PRXQuantum.3.030201,PhysRevB.90.064201} as well as being closely related to the geometry of spacetime \cite{PhysRevLett.96.181602, almheiri2021entropy, RevModPhys.90.035007}.  The evolution of entanglement in pure many-body quantum systems is concealed in the cooperative dynamics amongst the exponentially large number of modes of the system. Thus, except in rare cases, e.g., one-dimensional integrable models with conformal symmetry \cite{PasqualeCalabrese_2004, Calabrese_2007, HOLZHEY1994443}, obtaining a microscopic view of entanglement dynamics is generally not feasible in interacting many-body quantum systems. The most common approximation tools to explore many-body entanglement are variations of tensor-network-based numerical simulations, including Matrix Product States (MPS), which are mainly useful in the low-entanglement-density regimes \cite{verstraete2008matrix, ORUS2014117}. This paper will introduce a new means to study multipartite entanglement dynamics in many-body quantum systems, which we refer to as the Quantum Correlation Transfer Function (QCTF) formulation. Additionally, we will use the QCTF to provide a geometric interpretation of entanglement.

\par The QCTF provides a means to study correlation dynamics in many-body quantum systems. We exploit the fact that correlations between the constituents of a quantum system can also be obtained from the \textit{patterns} in its wave function or density matrix, rather than their elements. The investigation of such patterns is enabled through the employment of the Z-transformation, wherein the unitary time-evolution of a pure quantum system is mapped into a new complex-valued function, that constitutes the QCTF. As a result, the dynamics of entanglement is encoded in the analytic properties (e.g., poles, zeros, and residues) of the QCTF. In this general framework, various mathematical tools can be used to simplify the analysis. Most importantly, the QCTF permits bypassing the direct evaluation of the time evolution while nevertheless permitting the study of entanglement dynamics. 
\par Employing the QCTF, can either lead to a formulation of the quantum system which is fully dependent on the initially chosen basis, or alternatively, to a completely basis-independent (i.e., tensorial) description, where tensor indices represent the degrees of freedom in the Hilbert space. The basis-dependent approach may be implemented to study entanglement, given the Hamiltonian and the initial state of the system. In this case, by properly choosing the basis used in the Z-transformation, a closed-form complex function, or it's Laurent expansion, can be obtained upon making the QCTF transformation, from which entanglement dynamics is revealed by finding the QCTF residues. For example, this approach has been used to obtain the exact single-magnon entanglement dynamics in integrable Heisenberg chains \cite{magnon}.

\par  The basis-independent approach will be employed to present an extrinsic geometric portrayal of the QCTF-expressed entanglement (the second objective of the paper). In particular, based on the tensorial description of entanglement in the QCTF formulation, a \textit{natural} geometrical description follows. We show that the collective total squared areas spanned by the \textit{marginal} wave functions is a geometric measure of entanglement between parts of a pure quantum system. A few clarifications on the terminology above are called for. The word ``natural" reflects that this description mirrors the nature of entanglement between subsystems, which is, how different degrees of freedom in the subsystem evolve distinctly due to interaction with the remainder of the system. By ``marginal", we refer to the projection of the wave function onto an arbitrary basis of the vector space spanning  the subsystem of interest. This formulation can enable employing various tools from geometry to better understand entanglement. Since the focus of this paper is on constant Hamiltonians, with the formulation being $U(1)$ gauge-invariant, the geometric analysis is simply carried out on $\mathbb{C}^n$ with Euclidean metric (which highly simplifies the theory), and not the complex projective space, $\mathbb{C}\mathbf{P}^{n}$, with Fubini-Study metric \cite{kobayashi1963foundations, CAROLLO20201,PhysRevD.23.357}. We note that viewing quantum systems through the lens of geometry has often been fruitful in different areas of physics \cite{https://doi.org/10.1002/prop.201300020,penington2020entanglement,provost1980riemannian, PhysRevLett.99.100603,kibble1979geometrization,page1987geometrical,berry1984quantal,PhysRevLett.51.2167, kim2021information}, and it has led to links between the entropy of entanglement and the geometry of the space wherein the system's state lies \cite{PhysRevD.48.3743,THOOFT1985727, PatrickHayden_2007, strominger1996microscopic, PhysRevLett.96.181602,ShinseiRyu_2006, brody2001geometric, bengtsson2017geometry,wei2003geometric,grabowski2005geometry,man2017metric,FACCHI20104801,swingle2012geometric,Lambert_2023,vesperini2023entanglement,leinaas2006geometrical,ciaglia2022monotone}.

\par The paper is organized as follows. Section \ref{sub1b} introduces the transformation producing the QCTF. In Section \ref{sec:entanglement} a particular form of the QCTF transformation is presented for two-level subsystems along with its resultant entanglement measure formulation. Section \ref{geo} starts with a basic introduction to the geometry of the Hilbert space involved. Subsection \ref{geo:s1} gives the main results on the geometric structure of entanglement, and subsection \ref{renyii}, demonstrates the relation between the QCTF entanglement measure and the second-order R\'enyi entropy. Section \ref{concc} presents concluding remarks and suggestions for future research building on the foundation set out in the present work. The Appendices are referred to in the text for mathematical details.

\section{Basics of the QCTF Formulation} \label{sub1b}
\par The motivation behind the QCTF framework is to analyze a quantum system through the analytic behavior of a corresponding complex-valued function, referred to as QCTF. This function is obtained by employing the Z-transformation (initially proposed by Ragazzini and Zadeh to study sampled data \cite{6371274}) to either the entire or a portion of the wave function ($\ket{\tilde{\psi }(s)}=\mathcal{L}\{\ket{\psi (t)}\}$) in the Laplace domain; the density matrix (${\tilde{\rho }(s)}=\mathcal{L}\{{\rho (t)}\}$) may be analogously treated, as will be explained. The general case of transforming the wave function is given first to illustrate the use of the QCTF. We consider the unitary evolution of the initial state $\ket{\psi _0}$ governed by the time-independent Hamiltonian $\mathbf{H}$. Given any arbitrary basis set $\{\ket{j}| j=0: d-1\}$ that spans the Hilbert space associated with $\ket{\psi}$, then one form of the QCTF transformation can be defined as:
\begin{equation}\label{can}
\begin{split}
      \bar{\mathcal{K}} (z,s)&=\sum_{j=0}^{d-1}z^{j}\braket{j}{\tilde{\psi}(s)}
      \\&=\sum_{j=0}^{d-1}z^{j}\mel{j}{\mathbf{G}(s)}{\psi_0},
\end{split}
\end{equation} 
where $\mathbf{G}(s)=(s+\frac{i}{\hbar}\mathbf{H})^{-1}$ is the resolvent of $\mathbf{H}$ (the propagator of the Schr\"odinger's equation in the Laplace domain). The transformation (\ref{can}) implicitly depends on the basis $\{\ket{j}\}$ and how it is labeled (or numbered) due to the exponents $z^j$; nevertheless, as will be shown later, one can rewrite the QCTF fully independent of the choice of basis. We note that the transformation (\ref{can}) is an equivalent description of the quantum system's wave function, therefore the properties of the system are retained within the QCTF. The main purpose of this paper is to show that the entanglement evolution of subsystems can be effectively studied along with new insights via the QCTF formulation. 

\par The density matrix of a pure quantum system can be equivalently described in the QCTF formulation. The desired QCTF is a function of three complex variables, $z_d, z_a$ and $s$; Using (\ref{can}), it can be defined as

\begin{equation}
    \label{canII}
   {\mathcal{K}} (z_d,z_a,s)\doteq \bar{\mathcal{K}} (z_d z_a,s) \star \bar{\mathcal{K}}^* \big((z_a/ z_d)^*,s^*\big),
\end{equation}
where $\star$ denotes the ordinary product operation in the $z_d$ and $z_a$ domains along with the following convolution operation in the $s$ domain. If $\mathbf{F}_1(s)$ and $\mathbf{F}_2(s)$ are functions in the Laplace domain, then
\begin{equation}\label{conv}
    \mathbf{F}_1(s)\star \mathbf{F}_2(s)\doteq \frac{1}{2\pi i} \int_{-\infty}^{\infty} {\mathbf{F}_1(\sigma + i \omega) \mathbf{F}_2(s-\sigma -i \omega) d\omega},
\end{equation}
for some real $\sigma$ in the region of convergence of $\mathbf{F}_1 (s)$. This operation is the frequency-domain equivalent of the product of these functions in the time domain. In the important special case of simple poles, we have $(s+i\omega_1)^{-1}\star (s+i\omega_2)^{-1}=(s+i(\omega_1+\omega_2))^{-1}$. The QCTF (\ref{canII}) can be interpreted as a two-dimensional Z-transformation of the density matrix in the Laplace domain (see Appendix \ref{app1} for details).

\par More generally, the dynamics of a pure quantum system evolving under the Hamiltonian $\mathbf{H}$ can be represented by an operator function $\mathcal{H}(\mathbf{H},z_d,z_a,s)$, which includes its dynamical correlation properties, independent of its initial state. To see this feature, using (\ref{canII}), the QCTF transformation can be rewritten as the expectation value of $\mathcal{H}$ with respect to the arbitrary initial state ($\ket{\psi_0}$),
\begin{subequations} \label{70}
\begin{equation}
    \mathcal{K}(z_d,z_a,s)=\expval{\mathcal{H}(\mathbf{H},z_d,z_a,s)}{\psi_0},
\end{equation}
\begin{equation}
    \mathcal{H}={\textbf{G}^{\dag}(s^*)\bigg( \sum_{ i,j} {z_a}^{i+j} {z_d}^{i-j} \ket{j}{\bra{i}\bigg)\star \textbf{G}(s)}}.
\end{equation}
\end{subequations}
Operator $\mathcal{H}$ captures the dynamical properties of the quantum system, including the evolution of correlation between its constituent subsystems. This property, which will be discussed in detail in the next section, can be employed to understand how the correlation behavior varies for different initial states of the quantum system. 

Equation (\ref{70}) is the dual Laurent series expansions of $\mathcal{K}$, in both $z_d$ and $z_a$ variables, centered at the origins of the respective spaces. Based on the labeling $j=0,\cdots ,d-1$ for the Hilbert space basis, the origin $z_d=0$ is a pole of order (at most) $d-1$, while the origin $z_a=0$ is a removable singularity in the $z_a$ space. Also, for finite $d$, the function $\mathcal{K}$ is holomorphic in any punctured neighborhood of the origin, in both spaces. 

\par The QCTF (\ref{canII}) provides an equivalent description for the evolution of a quantum system in terms of its density matrix, $\rho (t)$. Therefore, desired matrix operations can be equivalently carried out on either $\mathcal{K}$ or functions of  $\mathcal{K}$ by a variety of means from complex analysis (e.g., finding residues, etc.). As an elementary example, given another arbitrary basis set $\{\ket{j'}\}$, the element $\mel{i'}{\tilde{\rho}}{j'}$ can be obtained by taking the inverse Laplace transform after utilizing the following residues (see Appendix \ref{app2} for the proof)
%\begin{widetext}
\begin{equation}\label{50}
\begin{split}
         \underset{{\substack{z_d=0\\z_a=0}}}{\mathbf{Res}} \Big (\sum_{i,j} \braket{j}{j'} \braket {i'}{i} z_d^{-(i-j)-1}z_a^{-(i+j)-1} {\mathcal{K}} (z_d,z_a,s)\Big ).
\end{split}
\end{equation}
%\end{widetext}
%where $\partial C_{a(d)}$ is any counter-clock-wise (CCW) closed curve in the $z_{a(d)}$ planes enclosing the origin, wherein $\mathcal{K}$ is holomorphic except at $z_{a(d)}=0$, and $\partial C_s$ is any CCW closed curve enclosing all of the poles of $\mathcal{K}$ on the imaginary axis of the $s$ domain.
\par To summarize, in the QCTF formulation, the goal is to study aspects of a quantum system from the \textit{patterns} in the elements of the system's state. These patterns reveal themselves through the encoding in the analytical properties of the QCTF (or functions of the QCTF) and can be found either by closed-contour integration or through the properties of the dual Laurent series expansion involved. The patterns arise through the Z-transformation's inherent summation over \textit{all} elements of the system's state. The main goal is to simplify the analysis of entanglement and express the results in a new perspective by either strategically choosing the QCTF basis labels or constructing a basis-independent formulation.

%\par In the QCTF formulation, the general approach to obtain the many-body quantum system's properties (e.g., the correlation dynamics between the particles) is to integrate a function of the QCTF (\ref{can}) around appropriate closed contours in the structural frequency spaces. Equation (\ref{50}) is an example of this procedure to obtain the system's density matrix elements. Given the fact that QCTF is an equivalent representation of the system's density matrix, in principle it is possible to obtain the dynamics of various forms of correlation, such as two-particle correlations, higher-order (multi-particle) correlations and particle's entanglement (the subject of the next section), by integrating a suitable function of the QCTF.

\section{Entanglement Dynamics for Two-level Subsystems}\label{sec:entanglement}
\par The application of the generic QCTF many-body framework in Section \ref{sub1b} is now focused on the entanglement dynamics of two-level subsystems when the overall system is initialized in a pure state.  Section \ref{geo} will build upon this scenario to consider entanglement dynamics for subsystems with more than two energy levels. We first introduce the entanglement measure used in the analysis and then its time evolution will be obtained from the system's QCTF. 

\par Here we consider a closed, discrete, bipartite quantum system, consisting of a two-level subsystem (referred to as subsystem $\mathcal{M}$) interacting with an accompanying $d$-dimensional quantum subsystem $\mathcal{R}$, that evolves according to the Hamiltonian $\mathbf{H}$ from the initial state $\ket{\psi_0}=\ket{\psi(t=0)}$. If we denote the reduced density matrix of the subsystem $\mathcal{M}$ by $\rho_{\mathcal{M}}(t)=\mathbf{\Tr}_{\mathcal{R}}\{\dyad{{\psi(t)}}\}$, then $\mathcal{Q}_{\mathcal{M}}(t)= \det (\rho_{\mathcal{M}}(t))$ is a time-dependent entanglement measure of subsystem $\mathcal{M}$, which is also monotonically related (i.e., they concurrently increase, decrease, or don't change with an infinitesimal change in the density matrix) to the second-order R\'enyi entanglement entropy through $ \mathcal{S}_2({\mathcal{M}})=-\log_2 (1-2\mathcal{Q_{M}})$. Equivalently, we will use its Laplace transformation $\tilde{\mathcal{Q}}_{\mathcal{M}}(s)=\mathcal{L}\{\mathcal{Q}_{\mathcal{M}}(t)\}$ as the dynamical entanglement measure in the analysis. 
\par The entanglement measure $\tilde{\mathcal{Q}}_{\mathcal{M}}(s)$ can be obtained from the QCTF. Given any basis for the quantum system which is constructed from the arbitrary basis vectors $\{\ket{+}, \ket{-}\}$ for $\mathcal{M}$ and $\{\ket{j},j=0,...,d-1\}$ ($d$ can be countably infinite) for $\mathcal{R}$, we define the QCTF on the off-diagonal block $\mel{+}{\tilde{\rho}(s)}{-}$ (which fully describes the bipartite entanglement between the subsystems) as,

\begin{subequations}\label{qqr}
\begin{equation}\label{generator}
 \mathcal{H}=\textbf{G}^{\dag}(s^*)\bigg( \sum_{i,j}{z_a}^{i+j}{z_d}^{i-j}\ket{-\otimes j}{\bra{ +\otimes i}\bigg)\star \textbf{G}(s)},
\end{equation}
\begin{equation}\label{QCTFE}
       \mathcal{K}(z_d,z_a,s)=\expval{\mathcal{H}}{\psi_0}.
\end{equation}
\end{subequations}

\par Having introduced the form of QCTF for the purpose of this section,  Appendix \ref{app3} proves that the dynamical entanglement measure ($\tilde{\mathcal{Q}}_{\mathcal{M}}(s)$) can be obtained from the QCTF (\ref{QCTFE}) as,

\begin{equation}\label{main}
\begin{split}
    \Tilde{\mathcal{Q}}_M(s)=&\underset{{\substack{z_d=0\\z_a=0}}}{\mathbf{Res}}\big((z_d z_a)^{-1}{\mathcal{K}(z_d,z_a,s)\star \mathcal{K}^*(1/z^*_d,1/z^*_a,s^*)} \big)\\&-{\mathcal{K}_d(s)\star\mathcal{K}_d^*(s^*)},
    \end{split}
\end{equation}
with $\mathcal{K}_d(s)=\eval{\underset{{\substack{z_d=0}}}{\mathbf{Res}}\big(z_d^{-1}{\mathcal{K}(z_d,z_a,s)}\big )}_{z_a=1}$, where $\underset{{{z=a}}}{\mathbf{Res}}(f(z))$ denotes the residue of $f$ at $z=a$. The first term in (\ref{main}) corresponds to the Frobenius norm of the off-diagonal sub-matrix $\mel{+}{\tilde{\rho}(s)}{-}$, while the second term corresponds to the summation of the cross-correlation of its diagonal. These two quantities are identical when the subsystems $\mathcal{M}$ and $\mathcal{R}$ are not entangled. In this formulation, direct evaluation of system's time evolution is avoided, and entanglement is directly obtained from the Hamiltonian's features and the initial wave function, which are reflected in (\ref{qqr}). Similar to the discussion in the previous section, the present QCTF formulation can depend on the choice of basis sets $\{j\}$ and $\{\ket{+}, \ket{-}\}$. Nevertheless, as will be shown shortly, upon finding the residues in (\ref{main}), one is left with a basis-independent expression that is geometrically meaningful. Concomitantly, strategically choosing the basis sets and their labels can simplify the QCTF formulation, similar to the theoretical analysis in \cite{https://doi.org/10.48550/arxiv.2201.11223, azodi2024emergencelightconeslongrange}.

\par Denoting $c_k\doteq \braket{k}{\psi_0}$ and the eigenmodes of the system's Hamiltonian by $\{\ket{k}, E_k\}$, the resolvent is
\begin{equation}
    \mathbf{G}(s)=\sum_k {(s+\frac{i}{\hbar}E_k)^{-1} \ket{k}\bra{k}}.
\end{equation}
Using the QCTF transformation (\ref{generator}), and then  (\ref{main}), the entanglement measure for the two-level subsystem of interest ($\mathcal{M}$), is simplified to:
\begin{equation}\label{enta}
    \Tilde{\mathcal{Q}}_M(s)=\sum_{k,l, k',l'} c_l c_{k}^* c_{k'} c_{l'}^*  \frac{ \braket{+k}{+k'} \braket{-l'}{-l}- \braket{+k}{-l} \braket{-l'}{+k'}}{ s+\frac{i}{\hbar}(E_{l}-E_{l'}-E_k+E_{k'})},
\end{equation}
where the inner products in the numerator, referred to as \textit{local overlaps}, are defined through the following decomposition:
\begin{equation}\label{deco}
\begin{split}
      \ket{k}=\ket{+k}\otimes\ket{+}+\ket{-k}\otimes\ket{-},
      \\ 
      (\text{equivalently}) \ket{\pm k}=\Big (\sum_{j} \big(\ket{j}(\bra{\pm}\otimes\bra{j})\big)\Big ) \ket{k}.
    %\\\ket{l}=\ket{+l}\otimes\ket{+}+\ket{-l}\otimes\ket{-}.
    \end{split}
\end{equation}
The vectors $\ket{\pm k}$ and $\ket{\pm l}$, which are not necessarily normalized, are in $\mathbb{C}^{{d}}$ and lie in the vector space underlying subsystem $\mathcal{R}$. Equation (\ref{enta}) involves three different components contributing to the entanglement dynamics. The first component is the product of the wave-function components multiplying the fraction, which describes the contribution from each eigenmode. The second component consists of the products of local overlaps in the numerator of the fraction, which is related to the geometric structure of the eigenstates and determines the strength of each eigenmode. The third component consists of the denominator of the fraction which determines the Laplace poles in the entanglement measure.
\par Equation (\ref{enta}) is a basis-invariant description of entanglement since the dependence on labels $i,j$ is no longer present. Also, the summation is independent of the choice of $\ket{\pm}$; to see this latter point, note that the first and third components (i.e., the coefficients and the denominator of the fraction) in equation (\ref{enta}) are invariant under the permutations $k \leftrightarrow l'$, $k' \leftrightarrow l$ and ($k \leftrightarrow l', k' \leftrightarrow l$). Using this symmetry, one can show the basis-invariance of (\ref{enta}) over the choice of $\ket{\pm}$ (by considering a unitary change of basis $\ket{\pm '}=U\ket{\pm }$). This observation enables using tensor algebra in the next section to give geometric insight into the entanglement measure (\ref{enta}). 
%\par Equations (\ref{generator}-\ref{QCTFE}, \ref{main}) give the entanglement dynamics through a dual Laurent series which can be used given , in contrast to the basis-independent formulation (\ref{enta}) that's a more abstract approach. In the next section, the geometric meaning of entanglement measure (\ref{enta}) is studied in an abstract setting.

%\par The first and third mechanisms in equation (\ref{enta}) are invariant under the following permutations of labels (symmetries) $k \leftrightarrow l'$ and $k' \leftrightarrow l$ and $k \leftrightarrow l', k' \leftrightarrow l$. Therefore, with enough care about equal indices, one can rewrite (\ref{enta}) with permuting indices for the term in the numerator (second mechanism).
%\begin{equation}\label{enta2}
%\begin{split}
%    \sum_{\substack{ k<l' \\ l<k'}} & c_l c_{k}^* c_{k'} c_{l'}^* \Big ({ s+\frac{i}{\hbar}(E_{l}-E_{l'}-E_k+E_{k'})}\Big )^{-1} \\ \Big (&\braket{+k}{+k'} \braket{-l'}{-l}- \braket{+k}{-l} \braket{-l'}{+k'}
%    \\&+ k \leftrightarrow l'+ k' \leftrightarrow l +k \leftrightarrow l', k' \leftrightarrow l\Big).
%\end{split}
%\end{equation}

%\tdplotsetmaincoords{100}{0}
\begin{figure}
    \centering
    \begin{tikzpicture}%[tdplot_main_coords]
	%%% Edit the following coordinate to change the shape of your
	%%% cuboid
    %\color{black!20!white}

      \physicalpendulum 

      \pgfsetfillopacity{1}
      \node  at (3,1.55) {$\mathbb{C}^d$};
    %\color{no default}
	%% Vanishing points for perspective handling
	\coordinate (P1) at (-37cm,-2.9cm); % left vanishing point (To pick)
	\coordinate (P2) at (8cm,12.5cm); % right vanishing point (To pick)

	%% (A1) and (A2) defines the 2 central points of the cuboid
	\coordinate (A1) at (1em,0cm); % central top point (To pick)
	\coordinate (A2) at (9em,-1cm); % central bottom point (To pick)

	%% (A3) to (A8) are computed given a unique parameter (or 2) .8
	% You can vary .8 from 0 to 1 to change perspective on left side
	\coordinate (A3) at ($(P1)!0.9!(A2)$); % To pick for perspective 
	\coordinate (A4) at ($(P1)!0.9!(A1)$);

	% You can vary .8 from 0 to 1 to change perspective on right side
	\coordinate (A7) at ($(P2)!0.9!(A2)$);
	\coordinate (A8) at ($(P2)!0.9!(A1)$);

	%% Automatically compute the last 2 points with intersections
	\coordinate (A5) at
	  (intersection cs: first line={(A8) -- (P1)},
			    second line={(A4) -- (P2)});
	\coordinate (A6) at
	  (intersection cs: first line={(A7) -- (P1)}, 
			    second line={(A3) -- (P2)});

	%%% Depending of what you want to display, you can comment/edit
	%%% the following lines

	%% Possibly draw back faces

	\fill[gray!90] (A2) -- (A3) -- (A6) -- (A7) -- cycle; % face 6
	\node at (barycentric cs:A2=1,A3=1,A6=1,A7=1) {$|\bra{\hat{j}_1 \psi} \wedge \bra{\hat{j}_2\psi}|$ };
	
	\fill[gray!65] (A3) -- (A4) -- (A5) -- (A6) -- cycle; % face 3
	\node at (barycentric cs:A3=1,A4=1,A5=1,A6=1) {$|\bra{ \hat{j}_3\psi} \wedge \bra{\hat{j}_2\psi}|$};
	
	\fill[gray!40] (A5) -- (A6) -- (A7) -- (A8) -- cycle; % face 4
	\node at (barycentric cs:A5=1,A6=1,A7=1,A8=1) {$|\bra{\hat{j}_3 \psi} \wedge \bra{\hat{j}_1\psi}|$};
	
	\draw[thick,->]  (A6)-- (A5) ;\node [above] at (A5) {$\ket{\hat{j}_1 \psi}$};
	\draw[thick,->] (A6)-- (A3) ;\node [below] at (A3) {$\ket{\hat{j}_2 \psi}$};
	\draw[thick,->](A6) -- (A7) ;\node [right] at (A7) {$\ket{\hat{j}_3 \psi}$};

	%% Possibly draw front faces

	% \fill[orange] (A1) -- (A8) -- (A7) -- (A2) -- cycle; % face 1
	% \node at (barycentric cs:A1=1,A8=1,A7=1,A2=1) {\tiny f1};
	
	%% Possibly draw front lines
%	\draw[thick] (A1) -- (A2);

	% Possibly draw points
	% (it can help you understand the cuboid structure)
	\foreach \i in {1,2,...,8}
	{
	  %\draw[fill=black] (A\i) circle (0.15em)
	    node[above right] {\tiny \i};
	}
	% \draw[fill=black] (P1) circle (0.1em) node[below] {\tiny p1};
	% \draw[fill=black] (P2) circle (0.1em) node[below] {\tiny p2};
\end{tikzpicture}
    \caption{The area spanned by pairs of marginal (or projected) wave functions is depicted. These vectors, given by (\ref{expansion}), belong to $\mathbb{C}^d$, which is the vector-space underlying subsystem $\mathcal{R}$. Equivalently, up to normalization, marginal wave-function $\ket{\hat{j} \psi}$ is equal to the post-measurement (i.e., the projected wave function of subsystem $\mathcal{M}$, after projective measurement of subsystem $\mathcal{M}$, in the $\{\ket{\hat{j}}\}$ basis) wave-function of subsystem $\mathcal{R}$. The total squared areas (i.e., the three shaded regions in the Figure) gives the entanglement between the subsystems.}
    \label{fig1}
\end{figure}
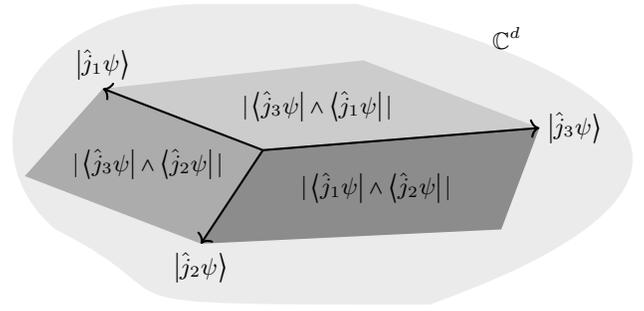

\par

\section{ Geometry of the QCTF entanglement}\label{geo}
In this section, a geometric interpretation of the QCTF entanglement measure (\ref{enta}) is presented. Toward this goal, we will start by introducing the following notation. Since the quantum system's Hilbert space is isomorphic to $\mathbb{C}^{2d}$, we will treat $\bra{z}$ as the (covariant) dual of $\ket{z}=\sum _{i} z^i \ket{i}$ with the Euclidean metric $g_{ij}=\delta_{ij}$, i.e. $\braket{z}{z}=\delta_{ij} z^i {z^*}^j=z^i z^*_i$, where in the last term we have used the Einstein summation convention. In this treatment, $\bra{z}$ is a \textit{1-form}.

\subsection{Exterior algebraic description of entanglement} \label{geo:s1}
\par Using the above terminology, the numerator of the fraction in the QCTF entanglement measure (\ref{enta}) can be written as a \textit{2-form} acting on the tensor product of two contravariant (ket) vectors, all of which belong to $\mathbb{C}^{d}$, as follows:
\begin{equation}\label{w1}
    \text{Numerator in (\ref{enta})}= (\bra{+k} \wedge \bra{-l'})(\ket{+k'}\otimes \ket{-l}),
\end{equation}
where the wedge operator ($\wedge$) is the alternating product, i.e., $\bra{z_1} \wedge \bra{z_2}=(\bra{z_1} \otimes \bra{z_2}-\bra{z_2} \otimes \bra{z_1})$. Note that in the remainder of the paper, any product between functions in the Laplace domain is understood to be carried out using the $\star$ operator defined in (\ref{conv}). Therefore, the \textit{evolution} of entanglement between $\mathcal{M}$ and $\mathcal{R}$ is given by the exterior product between the components of the Hamiltonian's eigenstates in directions corresponding to an orthonormal basis for subsystem $\mathcal{M}$. Building on (\ref{w1}), we can write the entire entanglement measure $\tilde{\mathcal {Q}}_{\mathcal{M}}(s)$ in (\ref{enta}) as:
\begin{equation}\label{w3}
    \tilde{\mathcal {Q}}_{\mathcal{M}}(s)=(\bra{+\psi} \wedge \bra{-\psi})(\ket{+\psi}\otimes \ket{-\psi}),
\end{equation}
where $\ket{\psi}= \ket{+\psi}\otimes \ket{+}+\ket{-\psi}\otimes \ket{-}$, which is $\ket{\pm\psi}=\Big (\sum_{j} \big(\ket{j}(\bra{\pm}\otimes\bra{j})\big)\Big ) \mathbf{G}(s)\ket{\psi_0}$. This description of entanglement is fully basis-independent (with respect to $\{\ket{\pm}\}$ and $\{\ket{j}\}$) and  $U(1)$ gauge invariant, i.e., $\ket{\psi}\rightarrow \ket{\tilde{\psi}}=e^{i\theta} \ket{\psi}$ for any real $\theta$. 
Geometrically, in the case of a two-level subsystem $\mathcal{M}$, the QCTF entanglement measures the squared area of the parallelogram spanned by $\ket{+\psi}$ and $\ket{-\psi}$ in  $\mathbb{C}^{{d}}$. 
\par This geometric observation can be extended to the QCTF entanglement measure for subsystem $\mathcal{M}$ with more than two energy levels. In this case, we denote an arbitrary basis for the sub-vector spaces underlying subsystem $\mathcal{M}$ ($n$-dimensional) and $\mathcal{R}$, respectively by $\{\ket{\hat{j}}; j=1,\cdots, n\}$ and $\{\ket{{i}}:i=1,\cdots,d\}$. Denoting the projected wavefunctions into its components along $\{\ket{\hat{j}}\}$ as  
\begin{equation}\label{expansion}
\begin{split}
      \ket{\hat{j} \psi}&=(\bra{\hat{j}}\otimes I_{\mathcal{R}}) \ket{\psi}\\&=\Big (\sum_{i} \big(\ket{i}(\bra{\hat{j}}\otimes\bra{i})\big)\Big ) \mathbf{G}(s)\ket{\psi_0}= \sum_{i} \psi_{\hat{j}{i}}\ket{i},
\end{split}  
\end{equation} 
the following expression in (\ref{qt}) is a measure of entanglement between subsystems $\mathcal{M}$ and $\mathcal{R}$ : The total collective squared areas spanned by pairs of projected wave functions (see Figure \ref{fig1}):
\begin{equation}\label{qt}
\begin{split}
    \tilde{\mathcal{Q}}_{\mathcal{M}}(s)=\sum_{1\leq \hat{j}_{1}<\hat{j}_2\leq n} {(\bra{\hat{j}_1 \psi} \wedge  \bra{\hat{j}_2 \psi})(\ket{\hat{j}_1 \psi} \otimes \ket{\hat{j}_2 \psi})}.  
\end{split}
\end{equation}
In the next subsection, it is shown that this entanglement measure is monotonically related to the second-order R\'enyi entropy. 
\par Given this extension of $\tilde{\mathcal{Q}}_{\mathcal{M}}(s)$ to subsystems with arbitrary energy levels, the QCTF transformation (\ref{qqr}) can be modified by considering an extra complex variable, $z_c$. Since the summation in (\ref{qt}) goes over pairs of basis vectors for subsystem $\mathcal{M}$, one can use an arbitrary labeling $h$ (i.e., by assigning a unique non-negative integer $h$ to each of the $\frac{n(n-1)}{2}$ pairs) for the set  $\{(\hat{j}_1,\hat{j}_2)| 1\le \hat{j}_1<\hat{j}_2\le n\}$ (we denote the pair labeled $h$ by $(h_{-}, h_{+})$). Therefore, the QCTF can be written as follows:

\begin{subequations}\label{qqre}
\begin{equation}
     \mathcal{H}=\textbf{G}^{\dag}(s^*)\bigg( \sum_{i,j,h}{z_a}^{i+j}{z_d}^{i-j}{z_c}^h\ket{h_{-}\otimes j}{\bra{ h_{+}\otimes i}\bigg)\star \textbf{G}(s)},
\end{equation}
\begin{equation}\label{QCTFEE}
       \mathcal{K}(z_d,z_a,z_c,s)=\expval{\mathcal{H}}{\psi_0}.
\end{equation}
\end{subequations}
Given this transformation, the entanglement measure $\Tilde{\mathcal{Q}}_M(s)$ is:
\begin{equation}\label{main22}
\begin{split}
    \Tilde{\mathcal{Q}}_M(s)=&\underset{{\substack{z_d=0\\z_a=0\\z_c=0}}}{\mathbf{Res}}\big(\frac{\mathcal{K}(z_d,z_a,z_c,s)\star \mathcal{K}^*(1/z^*_d,1/z^*_a,1/z^*_c,s^*)}{z_d z_a z_c} \big)\\&-\underset{{\substack{z_c=0}}}{\mathbf{Res}}\big (z_c^{-1}{\mathcal{K}_d(z_c,s)\star\mathcal{K}_d^*(1/z^*_c,s^*)}\big ),
    \end{split}
\end{equation}
with $\mathcal{K}_d(z_c,s)=\eval{\underset{{\substack{z_d=0}}}{\mathbf{Res}}\big(z_d^{-1}{\mathcal{K}(z_d,z_a,z_c,s)}\big )}_{z_a=1}$. By employing the variable $z_c$, and finding the residue (at the origin) in the last step, this process sums up the contribution from each pair of the projected wave functions.

%Here, the wedge (alternating) product is 
%\begin{equation}
%    \bra{M_{j_1} \psi} \wedge :\wedge \bra{M_{j_r} \psi}=\sum_{\pi} {\epsilon (\pi) \bra{M_{j_{\pi (1)}} \psi} \otimes  :\otimes \bra{M_{j_{\pi(r)}} \psi}},
%\end{equation}
%where $\pi$ denotes all possible permutations and $\epsilon(\pi)$ is the sign of the permutation. An interesting feature of this formula is that the summation runs up to forms of order $\min\{n,d \}$; the intuitive reason for this property is that the maximum order of correlations between the subsystems can not exceed the dimension of each subsystem. From the geometry point of view, as will be shown, this is a direct consequence of the following property of k-forms; in any $p-$ dimensional exterior product space of covariant vectors $\alpha_i$, for any $q>p$, we have $\alpha_1:\wedge:\alpha_q =0$.

\subsection{Relation with the second-order R\'enyi entropy} \label{renyii}
\par The entanglement measure (\ref{qt}) is monotonically related to the second-order R\'enyi entropy (${\mathbf{S}}^{(2)}= - \log_2 \big(\mathbf{\Tr({\tilde{\rho}}^2)}\big)$) of subsystem $\mathcal{M}$. To show this property, note that for each pair of $(j_1, j_2)$ in (\ref{qt}), the argument of the summation, ($(\bra{\hat{j{_1\psi}}}\wedge \bra{\hat{j}_2\psi})(\ket{\hat{j{_1\psi}}}\otimes \ket{\hat{j}_2\psi})$), is the collection of principal minors $$\begin{vmatrix}
\braket{\hat{j}_1\psi}{\hat{j}_1\psi} & \braket{\hat{j}_1\psi}{\hat{j}_2\psi}\\
\braket{\hat{j}_2\psi}{\hat{j}_1\psi} & \braket{\hat{j}_2\psi}{\hat{j}_2\psi}
\end{vmatrix} $$
of the reduced density matrix $\rho _{\mathcal{M}}$. Thus,
\begin{equation}
    \tilde{\mathcal{Q}}_{\mathcal{M}}(s)=\sum {\text{all $2\cross 2$ principal minors of $\rho _{\mathcal{M}}$ }}.
\end{equation}
Additionally, from matrix algebra, we have the following theorem \cite{meyer2023matrix}: The total sum of $2 \cross 2$ principal minors of any square matrix (here, the reduced density matrix $\rho _{\mathcal{M}}$) is equal to the second elementary symmetric polynomial of its eigenvalues $\{\lambda _i\}$:
\begin{equation}\label{esp}
    s_2=\sum_{1\le j_1< j_1\le n} \lambda_{j_1} \lambda _{j_2}.
\end{equation}
Thus,
\begin{equation}
    \tilde{\mathcal{Q}}_{\mathcal{M}}(s)=s_2=\frac{1}{2}(1-\mathbf{\Tr({\tilde{\rho}}^2)})=\frac{1}{2}(1-2^{-\mathbf{S}^{(2)})}).
\end{equation}
This equation relates the second-order R\'enyi entropy and the entanglement measure used in this paper. We should note that the quantity $1-\mathbf{\Tr({\tilde{\rho}}^2)}$ is sometimes referred to as the \textit{linear entropy} of $\mathcal{M}$ \cite{zyczkowski2003renyi}.

\section{Summary and Conclusion}\label{concc}
\par The new description of quantum many-body dynamics through the QCTF provides an equivalent representation of the wave-function or the density matrix that allows for obtaining entanglement dynamics in a quantum system \textit{directly} from its Hamiltonian. Therefore, the QCTF circumvents the bottleneck of evaluating the many-body system's evolution. The structure of the QCTF transformation enables using various forms of simplifications and algebraic manipulations. For example, the vector spaces' basis used in the QCTF transformation (denoted by $\{\ket{j}\}$ and $\{\ket{i'}\}$ in the paper) can be picked and labeled efficiently to give a closed-form function or a collection of them as sub-series of the transformation, e.g., each series might correspond to a perturbation order (similar to the application of QCTF in \cite{https://doi.org/10.48550/arxiv.2201.11223}). Another form of such simplifications offered by the QCTF is given in \cite{magnon}, wherein basis kets are labeled according to the single-magnon states that they are representing. 
\par Given the Hamiltonian and the initial state of the system, entanglement dynamics of subsystems can be obtained by finding the residues of the QCTF. Alternatively, the final QCTF entanglement can be described independently of the basis sets for each subsystem. In this case, entanglement was shown to have an exterior algebraic structure. More precisely, it is equal to the collective addition of norms of 2-forms constructed by the projected (i.e., marginal) wave functions. These norms quantify the squared area spanned by the corresponding marginal wave functions. 
\par  In this paper, a geometric portrayal of entanglement was presented as a measure of this quantity by its direct and equivalent consequences on the geometry of the system's state. This feature opens up the prospect of employing additional tools from geometry to understand and analyze entanglement in different fields and settings in future research. In particular, considering non-constant Hamiltonians (i.e., either with time-dependence or with Hamiltonians described by independent parameters) should lead to a \textit{differential geometric} description of the QCTF entanglement dynamics, which can be advantageous as it allows for using the generalized Stokes theorem (given the 2-differential form structure of the QCTF formulation) to study entanglement in these more complex scenarios. Additionally, the link between the QCTF entanglement formulation and the second-order R\'enyi entropy, discussed in subsection \ref{renyii}, which is established through a relation between the principal minors of the density matrix and the second elementary symmetric polynomials of its eigenvalues, leads to novel approaches to measure entanglement in experimental settings \cite{azodi2024measuringentanglementexploitingantisymmetric}. This opportunity arises because the elementary symmetric polynomials (\ref{esp}), fully describing the R\'enyi entropy, can equivalently be obtained in terms of local overlaps of the wavefunction, which are accessible experimentally \cite{garcia2013swap, huggins2021virtual}.

%\par In summary, this paper motivated the QCTF as a framework to fully capture the features of many-body dynamics
%\par In summary, this paper laid out a foundation for the QCTF formulation and its geometric description. Potential applications of the presented theoretical formulation include: \begin{enumerate}
    %obtaining dynamics of entanglement in many-body quantum systems, directly from the system Hamiltonian's properties. For example, in \cite{https://doi.org/10.48550/arxiv.2201.11223}, Many-Body Localized (MBL) entanglement dynamics of particles in disordered Heisenberg chain is studied, and in \cite{magnon} the exact entanglement dynamics is obtained in an integrable spin chain.
   %The geometric interpretation of entanglement and its Grassmannian algebraic formulation can potentially lead to new means of using tools from geometry and topology to understand many-body entanglement dynamics. In particular, by employing the Stokes–Cartan theorem and variants of the Gauss-Bonnet theorem (when considering non-constant Hamiltonians).
    %The QCTF formulation of entanglement can lead to novel means of measuring entanglement entropies in the experiment by relating the system's characteristics to observable quantities (as demonstrated in subsection \ref{renyii}).
%\end{enumerate}
%In this setting, finding links between the topology of the quantum system with subsystems' entanglement dynamics will be of high importance and priority. 

\begin{acknowledgments}
P.A acknowledges support from the Princeton Program in Plasma Science and Technology (PPST). H.R acknowledges support from the U.S Department Of Energy (DOE) grant (DE-FG02-02ER15344).
\end{acknowledgments}

%\appendix
\section*{Appendix}

\appendix \label{appee}

%\section{Prior Links between on Entanglement and Geometry} \label{introto}

\section{Derivation of equation (\ref{canII})}\label{app1}
Here, we show that the QCTF in (\ref{canII}) can be interpreted as a transformation of the system's density matrix by considering one chronological frequency, $s$, and two structural complex variables, $z_a$ and $z_d$. Using (\ref{can}) and the definition of operation $\star$, we can rewrite (\ref{canII}) as
\begin{equation}\label{BB1}
\begin{split}
     &{\mathcal{K}} (z_d,z_a,s)\doteq \bar{\mathcal{K}} (z_d z_a,s) \star \bar{\mathcal{K}}^* \big((z_a/ z_d)^*,s^*\big)\\&=\sum_{l,k=0}^{d-1}{\mel{l}{\mathbf{G}(s)}{\psi_0}\star \mel{\psi_0}{\mathbf{G}^\dag(s^*)}{k} z_d^{l-k}z_a^{l+k}}.    
\end{split}
\end{equation}
We define $\tilde{c}_l(s)\doteq\mel{l}{\mathbf{G}(s)}{\psi_0}=\mathcal{L}{\{\braket{l}{\psi (t)}\}}=\mathcal{L}{\{c_l(t)\}}$. Then using definition (\ref{conv}) we have:
\begin{widetext}
\begin{equation}\label{BB3}
\begin{split}
    \tilde{c}_l(s)\star \tilde{c}_k^*(s^*)&=\frac{1}{2\pi i} \int_{-\infty}^{\infty}{d\omega} {\tilde{c}_l(\sigma+i\omega)\tilde{c}_k^*\big((s-\sigma-i\omega)^*\big)}\\&=\int_{-\infty}^{\infty}{dt_1}\int_{-\infty}^{\infty}{dt_2}\frac{1}{2\pi i} \int_{-\infty}^{\infty}{d\omega}{c_l(t_1)c_k^*(t_2)e^{-st_2-\sigma(t_1-t_2)-i\omega(t_1-t_2)}}\\&=\int_{-\infty}^{\infty}{dt_1}\int_{-\infty}^{\infty}{dt_2}{c_l(t_1)c_k^*(t_2)e^{-st_2-\sigma(t_1-t_2)}\delta(t_1-t_2)}\\&=\int_{-\infty}^{\infty}{dt}{c_l(t)c_k^*(t)e^{-st}}=\mel{l}{\tilde{\rho}(s)}{k}.
\end{split}
\end{equation}
\end{widetext}
 
Therefore, substituting $\mel{l}{\mathbf{G}(s)}{\psi_0}\star \mel{\psi_0}{\mathbf{G}^\dag(s^*)}{k}$ with $\mel{l}{\tilde{\rho}(s)}{k}$ in (\ref{BB1}), gives:
\begin{equation}
         {\mathcal{K}} (z_d,z_a,s)=  \sum_{l,k=0}^{d-1}\mel{l}{\tilde{\rho}(s)}{k} z_d^{l-k}z_a^{l+k}.
\end{equation}

\section{Derivation of equation (\ref{50})}\label{app2}
Using equation (\ref{70}), we can rewrite the R.H.S of equation (\ref{50}) as follows
\begin{widetext}
\begin{equation}
     \frac{1}{(2\pi i)^3}\oint_{\partial C_s} \,ds \oint_{\partial C_d} \,dz_d \oint_{\partial C_a} \,dz_a e^{st'} \sum_{l,k} \sum _{l'',k''} \braket{k}{k'} \braket {l'}{l} z_d^{(l''-k'')-(l-k)-1}z_a^{(l''+k'')-(l+k)-1} \mel{l''}{\tilde{\rho}(s)}{k''}.
\end{equation}
\end{widetext}
Note that the three integrals are independent and can be evaluated interchangeably. We consider the $z_d$ and $z_a$ integrals first. Using the fact that the integrand is holomorphic inside the closed contours $\partial C_d$ and $\partial C_a$, except at the origin, by employing Cauchy's residue theorem the value of these two integrations are equal to the residue of the integrand at the origin $(z_d={0},z_a={0})$ times $(2\pi i)^2$. Also note that since $\mel{l''}{\tilde{\rho}(s)}{k''}$ has no $z_d, z_a$ dependence, the residue of the integrand in the $\mathbb{C}^2$ space is equal to the coefficient of $(z_az_d)^{-1}$ in the summation, which corresponds to the terms with $k''=k, l''=l$. Therefore, after evaluating the structural integrations, equation (\ref{50}) is
\begin{widetext}
\begin{equation}
    (\ref{50})=\frac{1}{2\pi i}\oint_{\partial C_s} \,ds e^{st'} \sum_{l,k} \braket {l'}{l}\mel{l}{\tilde{\rho}(s)}{k}\braket{k}{k'}=\mel{l'}{\frac{1}{2\pi i}\oint_{\partial C_s} \,ds e^{st'} \tilde{\rho}(s) }{k'}=\mel{l'}{\rho(t')}{k'},
\end{equation}
\end{widetext}
where we have used the definition of the inverse Laplace transform, and the linearity of the transform.

\section{Proof of equation (\ref{main})}\label{app3}
We start from the system's state in the time domain, $\ket{\psi(t)}$, and expand it in the product basis vectors of each subsystem, $\{\ket{+},\ket{-}\}$, and $\{\ket{l},l=0,...,{d-1}\}$, as follows
\begin{equation}
\begin{split}
        \ket{\psi(t)}=\sum_{a\in\{+,-\}}\sum_{l=0,...,{d-1}}{c_{al}(t){\ket{a\otimes l}}};\\\hspace {0cm}c_{al}(t)=\braket{a\otimes l}{{\psi(t)}}.
\end{split}
\end{equation}
Based on this expansion, we may construct the matrix $M_{2\times d}$ such that its first and second rows consist of $\{c_{+ l}(t)\}$ and $\{c_{- l}(t)\}$ respectively (the time dependence of the c's is not shown below for simplicity): 
\begin{equation}
M(t)\doteq    \begin{pmatrix}
c_{+ 0} & c_{+ 1} & \cdots & c_{+ l} & \cdots & c_{+ d-1}\\
c_{- 0} & c_{- 1} & \cdots & c_{- l} & \cdots & c_{- d-1}
\end{pmatrix}.
\end{equation}
If subsystem $\mathcal{M}$ is not entangled to subsystem $\mathcal{R}$ at $t=t'$, then $\ket{\psi(t')}$ is a product state and the rows of $M(t')$ are linearly dependent, i.e. $\rank \big(M(t')\big)=1$. This condition on the rank of $M$ is necessary and sufficient for the subsystem $\mathcal{M}$ to not be entangled to subsystem $\mathcal{R}$. Thus, if $rank(M)=2$ these subsystems are entangled. To construct a smooth indicator of entanglement, consider the following square sub-matrices $M^{ij}_{2\times 2}$, formed from the $i$th and $j$th columns of $M$,
\begin{equation}
    M^{ij}(t)\doteq
    \begin{pmatrix}
c_{+ i} & c_{+ j}  \\
c_{- i} & c_{- j} 
\end{pmatrix},
\end{equation}
and use them to define the entanglement measure $\mathcal{Q}_{\mathcal{M}}(t)$ as follows:
\begin{equation}\label{Q}
    \mathcal{Q}_{\mathcal{M}}(t)\doteq \sum_{0\leq i<j\leq d-1} {|\det \big(M^{ij}(t)\big)|^2}.
\end{equation}
More generally, $\mathcal{Q}_{\mathcal{M}}(t)=0$ is a necessary and sufficient condition for the subsystem $\mathcal{M}$ not to be entangled at $t$. 
\\ Expanding the summations in $\mathcal{Q}_(\mathcal{M}) (t)$ will result in
\begin{subequations}
\begin{align}
\mathcal{Q}_{\mathcal{M}}(t)=&\sum_{0\leq i<j\leq d-1}{\det(M^{ij})\big(\det(M^{ij})\big)^*}\\=&  \sum_{0\leq i<j\leq d-1}{( c_{+ i}c_{- j}-c_{+ j} c_{- i}) (c^*_{+ i}c^*_{- j}-c^*_{+ j} c^*_{- i})}\nonumber\\
    =&\sum_{0\leq i\neq j\leq d-1}{|c_{+ i}|^2|c_{- j}|^2}-\sum_{0\leq i\neq j\leq d-1}{c_{+ i}c_{- j} c^*_{+ j} c^*_{- i}} \nonumber\\
    =&\sum_{0\leq i, j\leq d-1}{|c_{+ i}|^2|c_{- j}|^2}-\sum_{0\leq i, j\leq d-1}{c_{+ i}c_{- j} c^*_{+ j} c^*_{- i}}\label{13c}\nonumber\\
    =&\big(\sum_{0\leq i\leq d-1}{|c_{+ i}|^2}\big) \big(\sum_{0\leq i\leq d-1}{|c_{- i}|^2}\big)\\-&\big(\sum_{0\leq i\leq d-1}{c_{+ i}c^*_{- i}}\big) \big(\sum_{0\leq i\leq d-1}{c^*_{+ i}c_{- i}}\big)\label{12d}\\
    =&\begin{vmatrix}
\sum_{i}{|c_{+ i}|^2} & \sum_{i}{\big(c_{+ i}c^*_{- i}\big)}\nonumber \\
\sum_{i}{\big(c^*_{+ i}c_{- i}\big)} & \sum_{i}{|c_{- i}|^2}
\end{vmatrix} 
=\det\big(\rho_{\mathcal{M}}(t)\big),  
\end{align}
\end{subequations}
which is the determinant of the time dependent reduced density matrix of subsystem $\mathcal{M}$. 
\par As the second step, here we prove that the R.H.S of equation (\ref{main}) is the Laplace transform of $\mathcal{Q}_\mathcal{M}(t)$. We start from the first term in  (\ref{main}). Before proceeding, we introduce the notation $\tilde{c}_{al}(s)\doteq \mathcal{L}\{c_{al}(t)\}=\mel{a\otimes l}{\mathbf{G}(s)}{{\psi_0}}$. Therefore, by using the definition of $\mathcal{K}(z_a,z_d,s)$ in (\ref{QCTFE}), the first term in (\ref{main}) and the fact that the $\star$ operation is associative and commutative , we have
\begin{widetext}
\begin{equation}\label{28}
   (2\pi i)^{-2}\sum_{l,k}\sum_{l',k'}\oint_{\partial C_d}\textbf{d}z_{d} \oint_{\partial C_a}\textbf{d}z_{a} {\tilde{c}_{+ l}(s)\star\tilde{c}^*_{- k}(s^*)\star{\tilde{c}^*_{+ l'}(s^*)\star\tilde{c}_{- k'}(s)z_a^{l+k-l'-k'-1}z_d^{l-k+k'-l'-1}}}.
\end{equation}
\end{widetext}
Upon employing Cauchy's residue theorem, based on the definition of the closed contours $\partial C_d$ and $\partial C_a$, then the value of the double integral is equal to $(2\pi i)^{2}$ times the residue of the integrand at the origin of the structural space $(z_d,z_a)$, which is the coefficient of $z_d^{-1}z_a^{-1}$ in the double summation. Therefore, the only remaining terms after the double integration must satisfy the following set of conditions:
\begin{equation}\label{con}
    \begin{cases}
               l+k-l'-k'-1=-1\\
               l-k+k'-l'-1=-1
            \end{cases}
            \iff
            \begin{cases}
               l=l'\\
               k=k'
            \end{cases}
.\end{equation}
Consequently, the first term in (\ref{main}) is:
\begin{widetext}

\begin{equation}
  \sum_{l,k} {\tilde{c}_{+ l}(s)\star{\tilde{c}^*_{+ l}(s^*)\star\tilde{c}_{- k}(s)}\star\tilde{c}^*_{- k}(s^*)}= \mathcal{L}\{\sum_{l, k}{|c_{+ l}|^2|c_{- k}|^2}\},
\end{equation}
\end{widetext}
which is the Laplace-dual of (\ref{13c}). In the last step, we operated with $\star$ similar to our calculation in equation (\ref{BB3}). Based on this equality, by showing that the second term in (\ref{main}) corresponds to (\ref{12d}), the proof is accomplished. To this end, we first consider $\mathcal{K}_d(z_a,s)$. Based on the definitions (\ref{QCTFE}) and (\ref{main}) we have:
\begin{equation}
    \mathcal{K}_d(z_a,s)= (2\pi i)^{-1} \sum_{l,k}\oint_{\partial C_d}\textbf{d}z_{d}{{\tilde{c}_{+ l}(s)\star {\tilde{c}^*_{- k}(s^*)z_a^{l+k}z_d^{l-k-1}}}}.
\end{equation}
Applying Cauchy's residue theorem, the integral is equal to $(2 \pi i)$ times the residue of the integrand at the origin of the $z_d$ space, which is the coefficient of $z_d^{-1}$ in it's Maclaurin series. Therefore, it is easy to see that for this integrand, only the terms with $l=k$ remain after the integration and we have
\begin{equation}
    \mathcal{K}_d(z_a,s)= \sum_{l}{\tilde{c}_{+ l}(s)\star \tilde{c}^*_{- l}(s^*)z_a^{2l}}.
\end{equation}
Consequently, the second term in (\ref{main}) is
\begin{widetext}

\begin{equation}
   \eval{\sum_{l,k}{\tilde{c}_{+ l}(s)\star {\tilde{c}^*_{- l}(s^*)\star \tilde{c}_{+ k}^*(s^*)\star {\tilde{c}_{- k}(s)}z_a^{2(l-k)}}}}_{z_a=1}=\mathcal{L}\{\sum_{l,k}{c_{+ l}c_{- k} c^*_{+k} c^*_{- l}\}},
\end{equation}
\end{widetext}
which is the Laplace-dual of the second term in (\ref{12d}). Therefore, using the linearity of the Laplace transformation, the R.H.S of (\ref{main}) is equal to the Laplace transformation of $\mathcal{Q}_{\mathcal{M}}(t)$, which proves the assertion (\ref{main}).

%\nocite{*}

\bibliography{apssamp}% Produces the bibliography via BibTeX.

%apsrev4-2.bst 2019-01-14 (MD) hand-edited version of apsrev4-1.bst
%Control: key (0)
%Control: author (8) initials jnrlst
%Control: editor formatted (1) identically to author
%Control: production of article title (0) allowed
%Control: page (0) single
%Control: year (1) truncated
%Control: production of eprint (0) enabled
\providecommand{\noopsort}[1]{}\providecommand{\singleletter}[1]{#1}%
\begin{thebibliography}{55}%
\makeatletter
\providecommand \@ifxundefined [1]{%
 \@ifx{#1\undefined}
}%
\providecommand \@ifnum [1]{%
 \ifnum #1\expandafter \@firstoftwo
 \else \expandafter \@secondoftwo
 \fi
}%
\providecommand \@ifx [1]{%
 \ifx #1\expandafter \@firstoftwo
 \else \expandafter \@secondoftwo
 \fi
}%
\providecommand \natexlab [1]{#1}%
\providecommand \enquote  [1]{``#1''}%
\providecommand \bibnamefont  [1]{#1}%
\providecommand \bibfnamefont [1]{#1}%
\providecommand \citenamefont [1]{#1}%
\providecommand \href@noop [0]{\@secondoftwo}%
\providecommand \href [0]{\begingroup \@sanitize@url \@href}%
\providecommand \@href[1]{\@@startlink{#1}\@@href}%
\providecommand \@@href[1]{\endgroup#1\@@endlink}%
\providecommand \@sanitize@url [0]{\catcode `\\12\catcode `\$12\catcode
  `\&12\catcode `\#12\catcode `\^12\catcode `\_12\catcode `\%12\relax}%
\providecommand \@@startlink[1]{}%
\providecommand \@@endlink[0]{}%
\providecommand \url  [0]{\begingroup\@sanitize@url \@url }%
\providecommand \@url [1]{\endgroup\@href {#1}{\urlprefix }}%
\providecommand \urlprefix  [0]{URL }%
\providecommand \Eprint [0]{\href }%
\providecommand \doibase [0]{https://doi.org/}%
\providecommand \selectlanguage [0]{\@gobble}%
\providecommand \bibinfo  [0]{\@secondoftwo}%
\providecommand \bibfield  [0]{\@secondoftwo}%
\providecommand \translation [1]{[#1]}%
\providecommand \BibitemOpen [0]{}%
\providecommand \bibitemStop [0]{}%
\providecommand \bibitemNoStop [0]{.\EOS\space}%
\providecommand \EOS [0]{\spacefactor3000\relax}%
\providecommand \BibitemShut  [1]{\csname bibitem#1\endcsname}%
\let\auto@bib@innerbib\@empty
%</preamble>
\bibitem [{\citenamefont {Steane}(1998)}]{steane1998quantum}%
  \BibitemOpen
  \bibfield  {author} {\bibinfo {author} {\bibfnamefont {A.}~\bibnamefont
  {Steane}},\ }\bibfield  {title} {\bibinfo {title} {Quantum computing},\
  }\href@noop {} {\bibfield  {journal} {\bibinfo  {journal} {Reports on
  Progress in Physics}\ }\textbf {\bibinfo {volume} {61}},\ \bibinfo {pages}
  {117} (\bibinfo {year} {1998})}\BibitemShut {NoStop}%
\bibitem [{\citenamefont {Plenio}\ and\ \citenamefont
  {Virmani}(2007)}]{plenio2007introduction}%
  \BibitemOpen
  \bibfield  {author} {\bibinfo {author} {\bibfnamefont {M.~B.}\ \bibnamefont
  {Plenio}}\ and\ \bibinfo {author} {\bibfnamefont {S.}~\bibnamefont
  {Virmani}},\ }\bibfield  {title} {\bibinfo {title} {An introduction to
  entanglement measures.},\ }\href@noop {} {\bibfield  {journal} {\bibinfo
  {journal} {Quantum Inf. Comput.}\ }\textbf {\bibinfo {volume} {7}},\ \bibinfo
  {pages} {1} (\bibinfo {year} {2007})}\BibitemShut {NoStop}%
\bibitem [{\citenamefont {Kitaev}\ and\ \citenamefont
  {Preskill}(2006)}]{PhysRevLett.96.110404}%
  \BibitemOpen
  \bibfield  {author} {\bibinfo {author} {\bibfnamefont {A.}~\bibnamefont
  {Kitaev}}\ and\ \bibinfo {author} {\bibfnamefont {J.}~\bibnamefont
  {Preskill}},\ }\bibfield  {title} {\bibinfo {title} {Topological entanglement
  entropy},\ }\href {https://doi.org/10.1103/PhysRevLett.96.110404} {\bibfield
  {journal} {\bibinfo  {journal} {Phys. Rev. Lett.}\ }\textbf {\bibinfo
  {volume} {96}},\ \bibinfo {pages} {110404} (\bibinfo {year}
  {2006})}\BibitemShut {NoStop}%
\bibitem [{\citenamefont {Horodecki}\ \emph {et~al.}(2009)\citenamefont
  {Horodecki}, \citenamefont {Horodecki}, \citenamefont {Horodecki},\ and\
  \citenamefont {Horodecki}}]{RevModPhys.81.865}%
  \BibitemOpen
  \bibfield  {author} {\bibinfo {author} {\bibfnamefont {R.}~\bibnamefont
  {Horodecki}}, \bibinfo {author} {\bibfnamefont {P.}~\bibnamefont
  {Horodecki}}, \bibinfo {author} {\bibfnamefont {M.}~\bibnamefont
  {Horodecki}},\ and\ \bibinfo {author} {\bibfnamefont {K.}~\bibnamefont
  {Horodecki}},\ }\bibfield  {title} {\bibinfo {title} {Quantum entanglement},\
  }\href {https://doi.org/10.1103/RevModPhys.81.865} {\bibfield  {journal}
  {\bibinfo  {journal} {Rev. Mod. Phys.}\ }\textbf {\bibinfo {volume} {81}},\
  \bibinfo {pages} {865} (\bibinfo {year} {2009})}\BibitemShut {NoStop}%
\bibitem [{\citenamefont {Kaufman}\ \emph {et~al.}(2016)\citenamefont
  {Kaufman}, \citenamefont {Tai}, \citenamefont {Lukin}, \citenamefont
  {Rispoli}, \citenamefont {Schittko}, \citenamefont {Preiss},\ and\
  \citenamefont {Greiner}}]{kaufman2016quantum}%
  \BibitemOpen
  \bibfield  {author} {\bibinfo {author} {\bibfnamefont {A.~M.}\ \bibnamefont
  {Kaufman}}, \bibinfo {author} {\bibfnamefont {M.~E.}\ \bibnamefont {Tai}},
  \bibinfo {author} {\bibfnamefont {A.}~\bibnamefont {Lukin}}, \bibinfo
  {author} {\bibfnamefont {M.}~\bibnamefont {Rispoli}}, \bibinfo {author}
  {\bibfnamefont {R.}~\bibnamefont {Schittko}}, \bibinfo {author}
  {\bibfnamefont {P.~M.}\ \bibnamefont {Preiss}},\ and\ \bibinfo {author}
  {\bibfnamefont {M.}~\bibnamefont {Greiner}},\ }\bibfield  {title} {\bibinfo
  {title} {Quantum thermalization through entanglement in an isolated many-body
  system},\ }\href@noop {} {\bibfield  {journal} {\bibinfo  {journal}
  {Science}\ }\textbf {\bibinfo {volume} {353}},\ \bibinfo {pages} {794}
  (\bibinfo {year} {2016})}\BibitemShut {NoStop}%
\bibitem [{\citenamefont {Abanin}\ \emph {et~al.}(2019)\citenamefont {Abanin},
  \citenamefont {Altman}, \citenamefont {Bloch},\ and\ \citenamefont
  {Serbyn}}]{RevModPhys.91.021001}%
  \BibitemOpen
  \bibfield  {author} {\bibinfo {author} {\bibfnamefont {D.~A.}\ \bibnamefont
  {Abanin}}, \bibinfo {author} {\bibfnamefont {E.}~\bibnamefont {Altman}},
  \bibinfo {author} {\bibfnamefont {I.}~\bibnamefont {Bloch}},\ and\ \bibinfo
  {author} {\bibfnamefont {M.}~\bibnamefont {Serbyn}},\ }\bibfield  {title}
  {\bibinfo {title} {Colloquium: Many-body localization, thermalization, and
  entanglement},\ }\href {https://doi.org/10.1103/RevModPhys.91.021001}
  {\bibfield  {journal} {\bibinfo  {journal} {Rev. Mod. Phys.}\ }\textbf
  {\bibinfo {volume} {91}},\ \bibinfo {pages} {021001} (\bibinfo {year}
  {2019})}\BibitemShut {NoStop}%
\bibitem [{\citenamefont {Eisert}\ \emph {et~al.}(2015)\citenamefont {Eisert},
  \citenamefont {Friesdorf},\ and\ \citenamefont
  {Gogolin}}]{eisert2015quantum}%
  \BibitemOpen
  \bibfield  {author} {\bibinfo {author} {\bibfnamefont {J.}~\bibnamefont
  {Eisert}}, \bibinfo {author} {\bibfnamefont {M.}~\bibnamefont {Friesdorf}},\
  and\ \bibinfo {author} {\bibfnamefont {C.}~\bibnamefont {Gogolin}},\
  }\bibfield  {title} {\bibinfo {title} {Quantum many-body systems out of
  equilibrium},\ }\href@noop {} {\bibfield  {journal} {\bibinfo  {journal}
  {Nature Physics}\ }\textbf {\bibinfo {volume} {11}},\ \bibinfo {pages} {124}
  (\bibinfo {year} {2015})}\BibitemShut {NoStop}%
\bibitem [{\citenamefont {Sala}\ \emph {et~al.}(2020)\citenamefont {Sala},
  \citenamefont {Rakovszky}, \citenamefont {Verresen}, \citenamefont {Knap},\
  and\ \citenamefont {Pollmann}}]{sala2020ergodicity}%
  \BibitemOpen
  \bibfield  {author} {\bibinfo {author} {\bibfnamefont {P.}~\bibnamefont
  {Sala}}, \bibinfo {author} {\bibfnamefont {T.}~\bibnamefont {Rakovszky}},
  \bibinfo {author} {\bibfnamefont {R.}~\bibnamefont {Verresen}}, \bibinfo
  {author} {\bibfnamefont {M.}~\bibnamefont {Knap}},\ and\ \bibinfo {author}
  {\bibfnamefont {F.}~\bibnamefont {Pollmann}},\ }\bibfield  {title} {\bibinfo
  {title} {Ergodicity breaking arising from hilbert space fragmentation in
  dipole-conserving hamiltonians},\ }\href@noop {} {\bibfield  {journal}
  {\bibinfo  {journal} {Physical Review X}\ }\textbf {\bibinfo {volume} {10}},\
  \bibinfo {pages} {011047} (\bibinfo {year} {2020})}\BibitemShut {NoStop}%
\bibitem [{\citenamefont {Bianchi}\ \emph {et~al.}(2022)\citenamefont
  {Bianchi}, \citenamefont {Hackl}, \citenamefont {Kieburg}, \citenamefont
  {Rigol},\ and\ \citenamefont {Vidmar}}]{PRXQuantum.3.030201}%
  \BibitemOpen
  \bibfield  {author} {\bibinfo {author} {\bibfnamefont {E.}~\bibnamefont
  {Bianchi}}, \bibinfo {author} {\bibfnamefont {L.}~\bibnamefont {Hackl}},
  \bibinfo {author} {\bibfnamefont {M.}~\bibnamefont {Kieburg}}, \bibinfo
  {author} {\bibfnamefont {M.}~\bibnamefont {Rigol}},\ and\ \bibinfo {author}
  {\bibfnamefont {L.}~\bibnamefont {Vidmar}},\ }\bibfield  {title} {\bibinfo
  {title} {Volume-law entanglement entropy of typical pure quantum states},\
  }\href {https://doi.org/10.1103/PRXQuantum.3.030201} {\bibfield  {journal}
  {\bibinfo  {journal} {PRX Quantum}\ }\textbf {\bibinfo {volume} {3}},\
  \bibinfo {pages} {030201} (\bibinfo {year} {2022})}\BibitemShut {NoStop}%
\bibitem [{\citenamefont {Nanduri}\ \emph {et~al.}(2014)\citenamefont
  {Nanduri}, \citenamefont {Kim},\ and\ \citenamefont
  {Huse}}]{PhysRevB.90.064201}%
  \BibitemOpen
  \bibfield  {author} {\bibinfo {author} {\bibfnamefont {A.}~\bibnamefont
  {Nanduri}}, \bibinfo {author} {\bibfnamefont {H.}~\bibnamefont {Kim}},\ and\
  \bibinfo {author} {\bibfnamefont {D.~A.}\ \bibnamefont {Huse}},\ }\bibfield
  {title} {\bibinfo {title} {Entanglement spreading in a many-body localized
  system},\ }\href {https://doi.org/10.1103/PhysRevB.90.064201} {\bibfield
  {journal} {\bibinfo  {journal} {Phys. Rev. B}\ }\textbf {\bibinfo {volume}
  {90}},\ \bibinfo {pages} {064201} (\bibinfo {year} {2014})}\BibitemShut
  {NoStop}%
\bibitem [{\citenamefont {Ryu}\ and\ \citenamefont
  {Takayanagi}(2006{\natexlab{a}})}]{PhysRevLett.96.181602}%
  \BibitemOpen
  \bibfield  {author} {\bibinfo {author} {\bibfnamefont {S.}~\bibnamefont
  {Ryu}}\ and\ \bibinfo {author} {\bibfnamefont {T.}~\bibnamefont
  {Takayanagi}},\ }\bibfield  {title} {\bibinfo {title} {Holographic derivation
  of entanglement entropy from the anti--de sitter space/conformal field theory
  correspondence},\ }\href {https://doi.org/10.1103/PhysRevLett.96.181602}
  {\bibfield  {journal} {\bibinfo  {journal} {Phys. Rev. Lett.}\ }\textbf
  {\bibinfo {volume} {96}},\ \bibinfo {pages} {181602} (\bibinfo {year}
  {2006}{\natexlab{a}})}\BibitemShut {NoStop}%
\bibitem [{\citenamefont {Almheiri}\ \emph {et~al.}(2021)\citenamefont
  {Almheiri}, \citenamefont {Hartman}, \citenamefont {Maldacena}, \citenamefont
  {Shaghoulian},\ and\ \citenamefont {Tajdini}}]{almheiri2021entropy}%
  \BibitemOpen
  \bibfield  {author} {\bibinfo {author} {\bibfnamefont {A.}~\bibnamefont
  {Almheiri}}, \bibinfo {author} {\bibfnamefont {T.}~\bibnamefont {Hartman}},
  \bibinfo {author} {\bibfnamefont {J.}~\bibnamefont {Maldacena}}, \bibinfo
  {author} {\bibfnamefont {E.}~\bibnamefont {Shaghoulian}},\ and\ \bibinfo
  {author} {\bibfnamefont {A.}~\bibnamefont {Tajdini}},\ }\bibfield  {title}
  {\bibinfo {title} {The entropy of hawking radiation},\ }\href@noop {}
  {\bibfield  {journal} {\bibinfo  {journal} {Reviews of Modern Physics}\
  }\textbf {\bibinfo {volume} {93}},\ \bibinfo {pages} {035002} (\bibinfo
  {year} {2021})}\BibitemShut {NoStop}%
\bibitem [{\citenamefont {Nishioka}(2018)}]{RevModPhys.90.035007}%
  \BibitemOpen
  \bibfield  {author} {\bibinfo {author} {\bibfnamefont {T.}~\bibnamefont
  {Nishioka}},\ }\bibfield  {title} {\bibinfo {title} {Entanglement entropy:
  Holography and renormalization group},\ }\href
  {https://doi.org/10.1103/RevModPhys.90.035007} {\bibfield  {journal}
  {\bibinfo  {journal} {Rev. Mod. Phys.}\ }\textbf {\bibinfo {volume} {90}},\
  \bibinfo {pages} {035007} (\bibinfo {year} {2018})}\BibitemShut {NoStop}%
\bibitem [{\citenamefont {Calabrese}\ and\ \citenamefont
  {Cardy}(2004)}]{PasqualeCalabrese_2004}%
  \BibitemOpen
  \bibfield  {author} {\bibinfo {author} {\bibfnamefont {P.}~\bibnamefont
  {Calabrese}}\ and\ \bibinfo {author} {\bibfnamefont {J.}~\bibnamefont
  {Cardy}},\ }\bibfield  {title} {\bibinfo {title} {Entanglement entropy and
  quantum field theory},\ }\href
  {https://doi.org/10.1088/1742-5468/2004/06/P06002} {\bibfield  {journal}
  {\bibinfo  {journal} {Journal of Statistical Mechanics: Theory and
  Experiment}\ }\textbf {\bibinfo {volume} {2004}},\ \bibinfo {pages} {P06002}
  (\bibinfo {year} {2004})}\BibitemShut {NoStop}%
\bibitem [{\citenamefont {Calabrese}\ and\ \citenamefont
  {Cardy}(2007)}]{Calabrese_2007}%
  \BibitemOpen
  \bibfield  {author} {\bibinfo {author} {\bibfnamefont {P.}~\bibnamefont
  {Calabrese}}\ and\ \bibinfo {author} {\bibfnamefont {J.}~\bibnamefont
  {Cardy}},\ }\bibfield  {title} {\bibinfo {title} {Entanglement and
  correlation functions following a local quench: a conformal field theory
  approach},\ }\href {https://doi.org/10.1088/1742-5468/2007/10/P10004}
  {\bibfield  {journal} {\bibinfo  {journal} {Journal of Statistical Mechanics:
  Theory and Experiment}\ }\textbf {\bibinfo {volume} {2007}},\ \bibinfo
  {pages} {P10004} (\bibinfo {year} {2007})}\BibitemShut {NoStop}%
\bibitem [{\citenamefont {Holzhey}\ \emph {et~al.}(1994)\citenamefont
  {Holzhey}, \citenamefont {Larsen},\ and\ \citenamefont
  {Wilczek}}]{HOLZHEY1994443}%
  \BibitemOpen
  \bibfield  {author} {\bibinfo {author} {\bibfnamefont {C.}~\bibnamefont
  {Holzhey}}, \bibinfo {author} {\bibfnamefont {F.}~\bibnamefont {Larsen}},\
  and\ \bibinfo {author} {\bibfnamefont {F.}~\bibnamefont {Wilczek}},\
  }\bibfield  {title} {\bibinfo {title} {Geometric and renormalized entropy in
  conformal field theory},\ }\href
  {https://doi.org/https://doi.org/10.1016/0550-3213(94)90402-2} {\bibfield
  {journal} {\bibinfo  {journal} {Nuclear Physics B}\ }\textbf {\bibinfo
  {volume} {424}},\ \bibinfo {pages} {443} (\bibinfo {year}
  {1994})}\BibitemShut {NoStop}%
\bibitem [{\citenamefont {Verstraete}\ \emph {et~al.}(2008)\citenamefont
  {Verstraete}, \citenamefont {Murg},\ and\ \citenamefont
  {Cirac}}]{verstraete2008matrix}%
  \BibitemOpen
  \bibfield  {author} {\bibinfo {author} {\bibfnamefont {F.}~\bibnamefont
  {Verstraete}}, \bibinfo {author} {\bibfnamefont {V.}~\bibnamefont {Murg}},\
  and\ \bibinfo {author} {\bibfnamefont {J.~I.}\ \bibnamefont {Cirac}},\
  }\bibfield  {title} {\bibinfo {title} {Matrix product states, projected
  entangled pair states, and variational renormalization group methods for
  quantum spin systems},\ }\href@noop {} {\bibfield  {journal} {\bibinfo
  {journal} {Advances in physics}\ }\textbf {\bibinfo {volume} {57}},\ \bibinfo
  {pages} {143} (\bibinfo {year} {2008})}\BibitemShut {NoStop}%
\bibitem [{\citenamefont {Orús}(2014)}]{ORUS2014117}%
  \BibitemOpen
  \bibfield  {author} {\bibinfo {author} {\bibfnamefont {R.}~\bibnamefont
  {Orús}},\ }\bibfield  {title} {\bibinfo {title} {A practical introduction to
  tensor networks: Matrix product states and projected entangled pair states},\
  }\href {https://doi.org/https://doi.org/10.1016/j.aop.2014.06.013} {\bibfield
   {journal} {\bibinfo  {journal} {Annals of Physics}\ }\textbf {\bibinfo
  {volume} {349}},\ \bibinfo {pages} {117} (\bibinfo {year}
  {2014})}\BibitemShut {NoStop}%
\bibitem [{\citenamefont {Azodi}\ and\ \citenamefont
  {Rabitz}(2024{\natexlab{a}})}]{magnon}%
  \BibitemOpen
  \bibfield  {author} {\bibinfo {author} {\bibfnamefont {P.}~\bibnamefont
  {Azodi}}\ and\ \bibinfo {author} {\bibfnamefont {H.~A.}\ \bibnamefont
  {Rabitz}},\ }\bibfield  {title} {\bibinfo {title} {Entanglement propagation
  in integrable heisenberg chains from a new lens},\ }\href
  {https://doi.org/10.1088/2399-6528/ad829a} {\bibfield  {journal} {\bibinfo
  {journal} {Journal of Physics Communications}\ }\textbf {\bibinfo {volume}
  {8}},\ \bibinfo {pages} {105002} (\bibinfo {year}
  {2024}{\natexlab{a}})}\BibitemShut {NoStop}%
\bibitem [{\citenamefont {Kobayashi}\ and\ \citenamefont
  {Nomizu}(1963)}]{kobayashi1963foundations}%
  \BibitemOpen
  \bibfield  {author} {\bibinfo {author} {\bibfnamefont {S.}~\bibnamefont
  {Kobayashi}}\ and\ \bibinfo {author} {\bibfnamefont {K.}~\bibnamefont
  {Nomizu}},\ }\href@noop {} {\emph {\bibinfo {title} {Foundations of
  differential geometry}}},\ Vol.~\bibinfo {volume} {1}\ (\bibinfo  {publisher}
  {New York, London},\ \bibinfo {year} {1963})\BibitemShut {NoStop}%
\bibitem [{\citenamefont {Carollo}\ \emph {et~al.}(2020)\citenamefont
  {Carollo}, \citenamefont {Valenti},\ and\ \citenamefont
  {Spagnolo}}]{CAROLLO20201}%
  \BibitemOpen
  \bibfield  {author} {\bibinfo {author} {\bibfnamefont {A.}~\bibnamefont
  {Carollo}}, \bibinfo {author} {\bibfnamefont {D.}~\bibnamefont {Valenti}},\
  and\ \bibinfo {author} {\bibfnamefont {B.}~\bibnamefont {Spagnolo}},\
  }\bibfield  {title} {\bibinfo {title} {Geometry of quantum phase
  transitions},\ }\href
  {https://doi.org/https://doi.org/10.1016/j.physrep.2019.11.002} {\bibfield
  {journal} {\bibinfo  {journal} {Physics Reports}\ }\textbf {\bibinfo {volume}
  {838}},\ \bibinfo {pages} {1} (\bibinfo {year} {2020})},\ \bibinfo {note}
  {geometry of quantum phase transitions}\BibitemShut {NoStop}%
\bibitem [{\citenamefont {Wootters}(1981)}]{PhysRevD.23.357}%
  \BibitemOpen
  \bibfield  {author} {\bibinfo {author} {\bibfnamefont {W.~K.}\ \bibnamefont
  {Wootters}},\ }\bibfield  {title} {\bibinfo {title} {Statistical distance and
  hilbert space},\ }\href {https://doi.org/10.1103/PhysRevD.23.357} {\bibfield
  {journal} {\bibinfo  {journal} {Phys. Rev. D}\ }\textbf {\bibinfo {volume}
  {23}},\ \bibinfo {pages} {357} (\bibinfo {year} {1981})}\BibitemShut
  {NoStop}%
\bibitem [{\citenamefont {Maldacena}\ and\ \citenamefont
  {Susskind}(2013)}]{https://doi.org/10.1002/prop.201300020}%
  \BibitemOpen
  \bibfield  {author} {\bibinfo {author} {\bibfnamefont {J.}~\bibnamefont
  {Maldacena}}\ and\ \bibinfo {author} {\bibfnamefont {L.}~\bibnamefont
  {Susskind}},\ }\bibfield  {title} {\bibinfo {title} {Cool horizons for
  entangled black holes},\ }\href
  {https://doi.org/https://doi.org/10.1002/prop.201300020} {\bibfield
  {journal} {\bibinfo  {journal} {Fortschritte der Physik}\ }\textbf {\bibinfo
  {volume} {61}},\ \bibinfo {pages} {781} (\bibinfo {year} {2013})}\BibitemShut
  {NoStop}%
\bibitem [{\citenamefont {Penington}(2020)}]{penington2020entanglement}%
  \BibitemOpen
  \bibfield  {author} {\bibinfo {author} {\bibfnamefont {G.}~\bibnamefont
  {Penington}},\ }\bibfield  {title} {\bibinfo {title} {Entanglement wedge
  reconstruction and the information paradox},\ }\href@noop {} {\bibfield
  {journal} {\bibinfo  {journal} {Journal of High Energy Physics}\ }\textbf
  {\bibinfo {volume} {2020}},\ \bibinfo {pages} {1} (\bibinfo {year}
  {2020})}\BibitemShut {NoStop}%
\bibitem [{\citenamefont {Provost}\ and\ \citenamefont
  {Vallee}(1980)}]{provost1980riemannian}%
  \BibitemOpen
  \bibfield  {author} {\bibinfo {author} {\bibfnamefont {J.}~\bibnamefont
  {Provost}}\ and\ \bibinfo {author} {\bibfnamefont {G.}~\bibnamefont
  {Vallee}},\ }\bibfield  {title} {\bibinfo {title} {Riemannian structure on
  manifolds of quantum states},\ }\href@noop {} {\bibfield  {journal} {\bibinfo
   {journal} {Communications in Mathematical Physics}\ }\textbf {\bibinfo
  {volume} {76}},\ \bibinfo {pages} {289} (\bibinfo {year} {1980})}\BibitemShut
  {NoStop}%
\bibitem [{\citenamefont {Zanardi}\ \emph {et~al.}(2007)\citenamefont
  {Zanardi}, \citenamefont {Giorda},\ and\ \citenamefont
  {Cozzini}}]{PhysRevLett.99.100603}%
  \BibitemOpen
  \bibfield  {author} {\bibinfo {author} {\bibfnamefont {P.}~\bibnamefont
  {Zanardi}}, \bibinfo {author} {\bibfnamefont {P.}~\bibnamefont {Giorda}},\
  and\ \bibinfo {author} {\bibfnamefont {M.}~\bibnamefont {Cozzini}},\
  }\bibfield  {title} {\bibinfo {title} {Information-theoretic differential
  geometry of quantum phase transitions},\ }\href
  {https://doi.org/10.1103/PhysRevLett.99.100603} {\bibfield  {journal}
  {\bibinfo  {journal} {Phys. Rev. Lett.}\ }\textbf {\bibinfo {volume} {99}},\
  \bibinfo {pages} {100603} (\bibinfo {year} {2007})}\BibitemShut {NoStop}%
\bibitem [{\citenamefont {Kibble}(1979)}]{kibble1979geometrization}%
  \BibitemOpen
  \bibfield  {author} {\bibinfo {author} {\bibfnamefont {T.~W.}\ \bibnamefont
  {Kibble}},\ }\bibfield  {title} {\bibinfo {title} {Geometrization of quantum
  mechanics},\ }\href@noop {} {\bibfield  {journal} {\bibinfo  {journal}
  {Communications in Mathematical Physics}\ }\textbf {\bibinfo {volume} {65}},\
  \bibinfo {pages} {189} (\bibinfo {year} {1979})}\BibitemShut {NoStop}%
\bibitem [{\citenamefont {Page}(1987)}]{page1987geometrical}%
  \BibitemOpen
  \bibfield  {author} {\bibinfo {author} {\bibfnamefont {D.~N.}\ \bibnamefont
  {Page}},\ }\bibfield  {title} {\bibinfo {title} {Geometrical description of
  berry's phase},\ }\href@noop {} {\bibfield  {journal} {\bibinfo  {journal}
  {Physical Review A}\ }\textbf {\bibinfo {volume} {36}},\ \bibinfo {pages}
  {3479} (\bibinfo {year} {1987})}\BibitemShut {NoStop}%
\bibitem [{\citenamefont {Berry}(1984)}]{berry1984quantal}%
  \BibitemOpen
  \bibfield  {author} {\bibinfo {author} {\bibfnamefont {M.~V.}\ \bibnamefont
  {Berry}},\ }\bibfield  {title} {\bibinfo {title} {Quantal phase factors
  accompanying adiabatic changes},\ }\href@noop {} {\bibfield  {journal}
  {\bibinfo  {journal} {Proceedings of the Royal Society of London. A.
  Mathematical and Physical Sciences}\ }\textbf {\bibinfo {volume} {392}},\
  \bibinfo {pages} {45} (\bibinfo {year} {1984})}\BibitemShut {NoStop}%
\bibitem [{\citenamefont {Simon}(1983)}]{PhysRevLett.51.2167}%
  \BibitemOpen
  \bibfield  {author} {\bibinfo {author} {\bibfnamefont {B.}~\bibnamefont
  {Simon}},\ }\bibfield  {title} {\bibinfo {title} {Holonomy, the quantum
  adiabatic theorem, and berry's phase},\ }\href
  {https://doi.org/10.1103/PhysRevLett.51.2167} {\bibfield  {journal} {\bibinfo
   {journal} {Phys. Rev. Lett.}\ }\textbf {\bibinfo {volume} {51}},\ \bibinfo
  {pages} {2167} (\bibinfo {year} {1983})}\BibitemShut {NoStop}%
\bibitem [{\citenamefont {Kim}(2021)}]{kim2021information}%
  \BibitemOpen
  \bibfield  {author} {\bibinfo {author} {\bibfnamefont {E.-j.}\ \bibnamefont
  {Kim}},\ }\bibfield  {title} {\bibinfo {title} {Information geometry,
  fluctuations, non-equilibrium thermodynamics, and geodesics in complex
  systems},\ }\href@noop {} {\bibfield  {journal} {\bibinfo  {journal}
  {Entropy}\ }\textbf {\bibinfo {volume} {23}},\ \bibinfo {pages} {1393}
  (\bibinfo {year} {2021})}\BibitemShut {NoStop}%
\bibitem [{\citenamefont {Susskind}\ \emph {et~al.}(1993)\citenamefont
  {Susskind}, \citenamefont {Thorlacius},\ and\ \citenamefont
  {Uglum}}]{PhysRevD.48.3743}%
  \BibitemOpen
  \bibfield  {author} {\bibinfo {author} {\bibfnamefont {L.}~\bibnamefont
  {Susskind}}, \bibinfo {author} {\bibfnamefont {L.}~\bibnamefont
  {Thorlacius}},\ and\ \bibinfo {author} {\bibfnamefont {J.}~\bibnamefont
  {Uglum}},\ }\bibfield  {title} {\bibinfo {title} {The stretched horizon and
  black hole complementarity},\ }\href
  {https://doi.org/10.1103/PhysRevD.48.3743} {\bibfield  {journal} {\bibinfo
  {journal} {Phys. Rev. D}\ }\textbf {\bibinfo {volume} {48}},\ \bibinfo
  {pages} {3743} (\bibinfo {year} {1993})}\BibitemShut {NoStop}%
\bibitem [{\citenamefont {{'t Hooft}}(1985)}]{THOOFT1985727}%
  \BibitemOpen
  \bibfield  {author} {\bibinfo {author} {\bibfnamefont {G.}~\bibnamefont {{'t
  Hooft}}},\ }\bibfield  {title} {\bibinfo {title} {On the quantum structure of
  a black hole},\ }\href
  {https://doi.org/https://doi.org/10.1016/0550-3213(85)90418-3} {\bibfield
  {journal} {\bibinfo  {journal} {Nuclear Physics B}\ }\textbf {\bibinfo
  {volume} {256}},\ \bibinfo {pages} {727} (\bibinfo {year}
  {1985})}\BibitemShut {NoStop}%
\bibitem [{\citenamefont {Hayden}\ and\ \citenamefont
  {Preskill}(2007)}]{PatrickHayden_2007}%
  \BibitemOpen
  \bibfield  {author} {\bibinfo {author} {\bibfnamefont {P.}~\bibnamefont
  {Hayden}}\ and\ \bibinfo {author} {\bibfnamefont {J.}~\bibnamefont
  {Preskill}},\ }\bibfield  {title} {\bibinfo {title} {Black holes as mirrors:
  quantum information in random subsystems},\ }\href
  {https://doi.org/10.1088/1126-6708/2007/09/120} {\bibfield  {journal}
  {\bibinfo  {journal} {Journal of High Energy Physics}\ }\textbf {\bibinfo
  {volume} {2007}},\ \bibinfo {pages} {120} (\bibinfo {year}
  {2007})}\BibitemShut {NoStop}%
\bibitem [{\citenamefont {Strominger}\ and\ \citenamefont
  {Vafa}(1996)}]{strominger1996microscopic}%
  \BibitemOpen
  \bibfield  {author} {\bibinfo {author} {\bibfnamefont {A.}~\bibnamefont
  {Strominger}}\ and\ \bibinfo {author} {\bibfnamefont {C.}~\bibnamefont
  {Vafa}},\ }\bibfield  {title} {\bibinfo {title} {Microscopic origin of the
  bekenstein-hawking entropy},\ }\href@noop {} {\bibfield  {journal} {\bibinfo
  {journal} {Physics Letters B}\ }\textbf {\bibinfo {volume} {379}},\ \bibinfo
  {pages} {99} (\bibinfo {year} {1996})}\BibitemShut {NoStop}%
\bibitem [{\citenamefont {Ryu}\ and\ \citenamefont
  {Takayanagi}(2006{\natexlab{b}})}]{ShinseiRyu_2006}%
  \BibitemOpen
  \bibfield  {author} {\bibinfo {author} {\bibfnamefont {S.}~\bibnamefont
  {Ryu}}\ and\ \bibinfo {author} {\bibfnamefont {T.}~\bibnamefont
  {Takayanagi}},\ }\bibfield  {title} {\bibinfo {title} {Aspects of holographic
  entanglement entropy},\ }\href
  {https://doi.org/10.1088/1126-6708/2006/08/045} {\bibfield  {journal}
  {\bibinfo  {journal} {Journal of High Energy Physics}\ }\textbf {\bibinfo
  {volume} {2006}},\ \bibinfo {pages} {045} (\bibinfo {year}
  {2006}{\natexlab{b}})}\BibitemShut {NoStop}%
\bibitem [{\citenamefont {Brody}\ and\ \citenamefont
  {Hughston}(2001)}]{brody2001geometric}%
  \BibitemOpen
  \bibfield  {author} {\bibinfo {author} {\bibfnamefont {D.~C.}\ \bibnamefont
  {Brody}}\ and\ \bibinfo {author} {\bibfnamefont {L.~P.}\ \bibnamefont
  {Hughston}},\ }\bibfield  {title} {\bibinfo {title} {Geometric quantum
  mechanics},\ }\href@noop {} {\bibfield  {journal} {\bibinfo  {journal}
  {Journal of geometry and physics}\ }\textbf {\bibinfo {volume} {38}},\
  \bibinfo {pages} {19} (\bibinfo {year} {2001})}\BibitemShut {NoStop}%
\bibitem [{\citenamefont {Bengtsson}\ and\ \citenamefont
  {{\.Z}yczkowski}(2017)}]{bengtsson2017geometry}%
  \BibitemOpen
  \bibfield  {author} {\bibinfo {author} {\bibfnamefont {I.}~\bibnamefont
  {Bengtsson}}\ and\ \bibinfo {author} {\bibfnamefont {K.}~\bibnamefont
  {{\.Z}yczkowski}},\ }\href@noop {} {\emph {\bibinfo {title} {Geometry of
  quantum states: an introduction to quantum entanglement}}}\ (\bibinfo
  {publisher} {Cambridge university press},\ \bibinfo {year}
  {2017})\BibitemShut {NoStop}%
\bibitem [{\citenamefont {Wei}\ and\ \citenamefont
  {Goldbart}(2003)}]{wei2003geometric}%
  \BibitemOpen
  \bibfield  {author} {\bibinfo {author} {\bibfnamefont {T.-C.}\ \bibnamefont
  {Wei}}\ and\ \bibinfo {author} {\bibfnamefont {P.~M.}\ \bibnamefont
  {Goldbart}},\ }\bibfield  {title} {\bibinfo {title} {Geometric measure of
  entanglement and applications to bipartite and multipartite quantum states},\
  }\href@noop {} {\bibfield  {journal} {\bibinfo  {journal} {Physical Review
  A}\ }\textbf {\bibinfo {volume} {68}},\ \bibinfo {pages} {042307} (\bibinfo
  {year} {2003})}\BibitemShut {NoStop}%
\bibitem [{\citenamefont {Grabowski}\ \emph {et~al.}(2005)\citenamefont
  {Grabowski}, \citenamefont {Ku{\'s}},\ and\ \citenamefont
  {Marmo}}]{grabowski2005geometry}%
  \BibitemOpen
  \bibfield  {author} {\bibinfo {author} {\bibfnamefont {J.}~\bibnamefont
  {Grabowski}}, \bibinfo {author} {\bibfnamefont {M.}~\bibnamefont {Ku{\'s}}},\
  and\ \bibinfo {author} {\bibfnamefont {G.}~\bibnamefont {Marmo}},\ }\bibfield
   {title} {\bibinfo {title} {Geometry of quantum systems: density states and
  entanglement},\ }\href@noop {} {\bibfield  {journal} {\bibinfo  {journal}
  {Journal of Physics A: Mathematical and General}\ }\textbf {\bibinfo {volume}
  {38}},\ \bibinfo {pages} {10217} (\bibinfo {year} {2005})}\BibitemShut
  {NoStop}%
\bibitem [{\citenamefont {Man’ko}\ \emph {et~al.}(2017)\citenamefont
  {Man’ko}, \citenamefont {Marmo}, \citenamefont {Ventriglia},\ and\
  \citenamefont {Vitale}}]{man2017metric}%
  \BibitemOpen
  \bibfield  {author} {\bibinfo {author} {\bibfnamefont {V.~I.}\ \bibnamefont
  {Man’ko}}, \bibinfo {author} {\bibfnamefont {G.}~\bibnamefont {Marmo}},
  \bibinfo {author} {\bibfnamefont {F.}~\bibnamefont {Ventriglia}},\ and\
  \bibinfo {author} {\bibfnamefont {P.}~\bibnamefont {Vitale}},\ }\bibfield
  {title} {\bibinfo {title} {Metric on the space of quantum states from
  relative entropy. tomographic reconstruction},\ }\href@noop {} {\bibfield
  {journal} {\bibinfo  {journal} {Journal of Physics A: Mathematical and
  Theoretical}\ }\textbf {\bibinfo {volume} {50}},\ \bibinfo {pages} {335302}
  (\bibinfo {year} {2017})}\BibitemShut {NoStop}%
\bibitem [{\citenamefont {Facchi}\ \emph {et~al.}(2010)\citenamefont {Facchi},
  \citenamefont {Kulkarni}, \citenamefont {Man'ko}, \citenamefont {Marmo},
  \citenamefont {Sudarshan},\ and\ \citenamefont
  {Ventriglia}}]{FACCHI20104801}%
  \BibitemOpen
  \bibfield  {author} {\bibinfo {author} {\bibfnamefont {P.}~\bibnamefont
  {Facchi}}, \bibinfo {author} {\bibfnamefont {R.}~\bibnamefont {Kulkarni}},
  \bibinfo {author} {\bibfnamefont {V.}~\bibnamefont {Man'ko}}, \bibinfo
  {author} {\bibfnamefont {G.}~\bibnamefont {Marmo}}, \bibinfo {author}
  {\bibfnamefont {E.}~\bibnamefont {Sudarshan}},\ and\ \bibinfo {author}
  {\bibfnamefont {F.}~\bibnamefont {Ventriglia}},\ }\bibfield  {title}
  {\bibinfo {title} {Classical and quantum fisher information in the
  geometrical formulation of quantum mechanics},\ }\href
  {https://doi.org/https://doi.org/10.1016/j.physleta.2010.10.005} {\bibfield
  {journal} {\bibinfo  {journal} {Physics Letters A}\ }\textbf {\bibinfo
  {volume} {374}},\ \bibinfo {pages} {4801} (\bibinfo {year}
  {2010})}\BibitemShut {NoStop}%
\bibitem [{\citenamefont {Swingle}\ and\ \citenamefont
  {Senthil}(2012)}]{swingle2012geometric}%
  \BibitemOpen
  \bibfield  {author} {\bibinfo {author} {\bibfnamefont {B.}~\bibnamefont
  {Swingle}}\ and\ \bibinfo {author} {\bibfnamefont {T.}~\bibnamefont
  {Senthil}},\ }\bibfield  {title} {\bibinfo {title} {Geometric proof of the
  equality between entanglement and edge spectra},\ }\href@noop {} {\bibfield
  {journal} {\bibinfo  {journal} {Physical Review B}\ }\textbf {\bibinfo
  {volume} {86}},\ \bibinfo {pages} {045117} (\bibinfo {year}
  {2012})}\BibitemShut {NoStop}%
\bibitem [{\citenamefont {Lambert}\ and\ \citenamefont
  {Sørensen}(2023)}]{Lambert_2023}%
  \BibitemOpen
  \bibfield  {author} {\bibinfo {author} {\bibfnamefont {J.}~\bibnamefont
  {Lambert}}\ and\ \bibinfo {author} {\bibfnamefont {E.~S.}\ \bibnamefont
  {Sørensen}},\ }\bibfield  {title} {\bibinfo {title} {From classical to
  quantum information geometry: a guide for physicists},\ }\href
  {https://doi.org/10.1088/1367-2630/aceb14} {\bibfield  {journal} {\bibinfo
  {journal} {New Journal of Physics}\ }\textbf {\bibinfo {volume} {25}},\
  \bibinfo {pages} {081201} (\bibinfo {year} {2023})}\BibitemShut {NoStop}%
\bibitem [{\citenamefont {Vesperini}\ \emph {et~al.}(2023)\citenamefont
  {Vesperini}, \citenamefont {Bel-Hadj-Aissa},\ and\ \citenamefont
  {Franzosi}}]{vesperini2023entanglement}%
  \BibitemOpen
  \bibfield  {author} {\bibinfo {author} {\bibfnamefont {A.}~\bibnamefont
  {Vesperini}}, \bibinfo {author} {\bibfnamefont {G.}~\bibnamefont
  {Bel-Hadj-Aissa}},\ and\ \bibinfo {author} {\bibfnamefont {R.}~\bibnamefont
  {Franzosi}},\ }\bibfield  {title} {\bibinfo {title} {Entanglement and quantum
  correlation measures for quantum multipartite mixed states},\ }\href@noop {}
  {\bibfield  {journal} {\bibinfo  {journal} {Scientific Reports}\ }\textbf
  {\bibinfo {volume} {13}},\ \bibinfo {pages} {2852} (\bibinfo {year}
  {2023})}\BibitemShut {NoStop}%
\bibitem [{\citenamefont {Leinaas}\ \emph {et~al.}(2006)\citenamefont
  {Leinaas}, \citenamefont {Myrheim},\ and\ \citenamefont
  {Ovrum}}]{leinaas2006geometrical}%
  \BibitemOpen
  \bibfield  {author} {\bibinfo {author} {\bibfnamefont {J.~M.}\ \bibnamefont
  {Leinaas}}, \bibinfo {author} {\bibfnamefont {J.}~\bibnamefont {Myrheim}},\
  and\ \bibinfo {author} {\bibfnamefont {E.}~\bibnamefont {Ovrum}},\ }\bibfield
   {title} {\bibinfo {title} {Geometrical aspects of entanglement},\
  }\href@noop {} {\bibfield  {journal} {\bibinfo  {journal} {Physical Review
  A}\ }\textbf {\bibinfo {volume} {74}},\ \bibinfo {pages} {012313} (\bibinfo
  {year} {2006})}\BibitemShut {NoStop}%
\bibitem [{\citenamefont {Ciaglia}\ \emph {et~al.}(2022)\citenamefont
  {Ciaglia}, \citenamefont {Di~Cosmo}, \citenamefont {Di~Nocera},\ and\
  \citenamefont {Vitale}}]{ciaglia2022monotone}%
  \BibitemOpen
  \bibfield  {author} {\bibinfo {author} {\bibfnamefont {F.~M.}\ \bibnamefont
  {Ciaglia}}, \bibinfo {author} {\bibfnamefont {F.}~\bibnamefont {Di~Cosmo}},
  \bibinfo {author} {\bibfnamefont {F.}~\bibnamefont {Di~Nocera}},\ and\
  \bibinfo {author} {\bibfnamefont {P.}~\bibnamefont {Vitale}},\ }\bibfield
  {title} {\bibinfo {title} {Monotone metric tensors in quantum information
  geometry},\ }\href@noop {} {\bibfield  {journal} {\bibinfo  {journal} {arXiv
  preprint arXiv:2203.10857}\ } (\bibinfo {year} {2022})}\BibitemShut {NoStop}%
\bibitem [{\citenamefont {Ragazzini}\ and\ \citenamefont
  {Zadeh}(1952)}]{6371274}%
  \BibitemOpen
  \bibfield  {author} {\bibinfo {author} {\bibfnamefont {J.~R.}\ \bibnamefont
  {Ragazzini}}\ and\ \bibinfo {author} {\bibfnamefont {L.~A.}\ \bibnamefont
  {Zadeh}},\ }\bibfield  {title} {\bibinfo {title} {The analysis of
  sampled-data systems},\ }\href {https://doi.org/10.1109/TAI.1952.6371274}
  {\bibfield  {journal} {\bibinfo  {journal} {Transactions of the American
  Institute of Electrical Engineers, Part II: Applications and Industry}\
  }\textbf {\bibinfo {volume} {71}},\ \bibinfo {pages} {225} (\bibinfo {year}
  {1952})}\BibitemShut {NoStop}%
\bibitem [{\citenamefont {Azodi}\ and\ \citenamefont
  {Rabitz}(2025)}]{https://doi.org/10.48550/arxiv.2201.11223}%
  \BibitemOpen
  \bibfield  {author} {\bibinfo {author} {\bibfnamefont {P.}~\bibnamefont
  {Azodi}}\ and\ \bibinfo {author} {\bibfnamefont {H.~A.}\ \bibnamefont
  {Rabitz}},\ }\href {https://arxiv.org/abs/2201.11223} {\bibinfo {title}
  {Stability and quasi-periodicity of many-body localized dynamics}} (\bibinfo
  {year} {2025}),\ \Eprint {https://arxiv.org/abs/2201.11223} {arXiv:2201.11223
  [quant-ph]} \BibitemShut {NoStop}%
\bibitem [{\citenamefont {Azodi}\ and\ \citenamefont
  {Rabitz}(2024{\natexlab{b}})}]{azodi2024emergencelightconeslongrange}%
  \BibitemOpen
  \bibfield  {author} {\bibinfo {author} {\bibfnamefont {P.}~\bibnamefont
  {Azodi}}\ and\ \bibinfo {author} {\bibfnamefont {H.~A.}\ \bibnamefont
  {Rabitz}},\ }\href {https://arxiv.org/abs/2407.11639} {\bibinfo {title}
  {Emergence of light cones in long-range interacting spin chains is due to
  destructive interference}} (\bibinfo {year} {2024}{\natexlab{b}}),\ \Eprint
  {https://arxiv.org/abs/2407.11639} {arXiv:2407.11639 [quant-ph]} \BibitemShut
  {NoStop}%
\bibitem [{\citenamefont {Meyer}(2023)}]{meyer2023matrix}%
  \BibitemOpen
  \bibfield  {author} {\bibinfo {author} {\bibfnamefont {C.~D.}\ \bibnamefont
  {Meyer}},\ }\href@noop {} {\emph {\bibinfo {title} {Matrix analysis and
  applied linear algebra}}},\ Vol.\ \bibinfo {volume} {188}\ (\bibinfo
  {publisher} {Siam},\ \bibinfo {year} {2023})\BibitemShut {NoStop}%
\bibitem [{\citenamefont {{\.Z}yczkowski}(2003)}]{zyczkowski2003renyi}%
  \BibitemOpen
  \bibfield  {author} {\bibinfo {author} {\bibfnamefont {K.}~\bibnamefont
  {{\.Z}yczkowski}},\ }\bibfield  {title} {\bibinfo {title} {R{\'e}nyi
  extrapolation of shannon entropy},\ }\href@noop {} {\bibfield  {journal}
  {\bibinfo  {journal} {Open Systems \& Information Dynamics}\ }\textbf
  {\bibinfo {volume} {10}},\ \bibinfo {pages} {297} (\bibinfo {year}
  {2003})}\BibitemShut {NoStop}%
\bibitem [{\citenamefont {Azodi}\ \emph {et~al.}(2024)\citenamefont {Azodi},
  \citenamefont {Lienhard},\ and\ \citenamefont
  {Rabitz}}]{azodi2024measuringentanglementexploitingantisymmetric}%
  \BibitemOpen
  \bibfield  {author} {\bibinfo {author} {\bibfnamefont {P.}~\bibnamefont
  {Azodi}}, \bibinfo {author} {\bibfnamefont {B.}~\bibnamefont {Lienhard}},\
  and\ \bibinfo {author} {\bibfnamefont {H.~A.}\ \bibnamefont {Rabitz}},\
  }\href {https://arxiv.org/abs/2409.17236} {\bibinfo {title} {Measuring
  entanglement by exploiting its anti-symmetric nature}} (\bibinfo {year}
  {2024}),\ \Eprint {https://arxiv.org/abs/2409.17236} {arXiv:2409.17236
  [quant-ph]} \BibitemShut {NoStop}%
\bibitem [{\citenamefont {Garcia-Escartin}\ and\ \citenamefont
  {Chamorro-Posada}(2013)}]{garcia2013swap}%
  \BibitemOpen
  \bibfield  {author} {\bibinfo {author} {\bibfnamefont {J.~C.}\ \bibnamefont
  {Garcia-Escartin}}\ and\ \bibinfo {author} {\bibfnamefont {P.}~\bibnamefont
  {Chamorro-Posada}},\ }\bibfield  {title} {\bibinfo {title} {Swap test and
  hong-ou-mandel effect are equivalent},\ }\href@noop {} {\bibfield  {journal}
  {\bibinfo  {journal} {Physical Review A}\ }\textbf {\bibinfo {volume} {87}},\
  \bibinfo {pages} {052330} (\bibinfo {year} {2013})}\BibitemShut {NoStop}%
\bibitem [{\citenamefont {Huggins}\ \emph {et~al.}(2021)\citenamefont
  {Huggins}, \citenamefont {McArdle}, \citenamefont {O’Brien}, \citenamefont
  {Lee}, \citenamefont {Rubin}, \citenamefont {Boixo}, \citenamefont {Whaley},
  \citenamefont {Babbush},\ and\ \citenamefont {McClean}}]{huggins2021virtual}%
  \BibitemOpen
  \bibfield  {author} {\bibinfo {author} {\bibfnamefont {W.~J.}\ \bibnamefont
  {Huggins}}, \bibinfo {author} {\bibfnamefont {S.}~\bibnamefont {McArdle}},
  \bibinfo {author} {\bibfnamefont {T.~E.}\ \bibnamefont {O’Brien}}, \bibinfo
  {author} {\bibfnamefont {J.}~\bibnamefont {Lee}}, \bibinfo {author}
  {\bibfnamefont {N.~C.}\ \bibnamefont {Rubin}}, \bibinfo {author}
  {\bibfnamefont {S.}~\bibnamefont {Boixo}}, \bibinfo {author} {\bibfnamefont
  {K.~B.}\ \bibnamefont {Whaley}}, \bibinfo {author} {\bibfnamefont
  {R.}~\bibnamefont {Babbush}},\ and\ \bibinfo {author} {\bibfnamefont {J.~R.}\
  \bibnamefont {McClean}},\ }\bibfield  {title} {\bibinfo {title} {Virtual
  distillation for quantum error mitigation},\ }\href@noop {} {\bibfield
  {journal} {\bibinfo  {journal} {Physical Review X}\ }\textbf {\bibinfo
  {volume} {11}},\ \bibinfo {pages} {041036} (\bibinfo {year}
  {2021})}\BibitemShut {NoStop}%
\end{thebibliography}%

\end{document}